\documentclass[aps,twocolumn,pra,amsmath,amssymb]{revtex4}
\usepackage{amssymb}
\usepackage{graphicx}

\def\be{\begin{equation}}
\def\ee{\end{equation}}
\def\bea{\begin{eqnarray}}
\def\eea{\end{eqnarray}}

\def\one{\leavevmode\hbox{\small1\normalsize\kern-.33em1}}

\def\compl{\begin{picture}(8,8)\put(0,0){C}\put(3,0.3){\line(0,1){7}}\end{picture}}
\newcommand{\w}[1]{(\ref{#1})}

\newcommand{\bra}[1]{\mbox{$\langle #1 |$}}
\newcommand{\ket}[1]{\mbox{$| #1 \rangle$}}

\def\tr{\mbox{tr}}

\begin{document}

\title{Key distillation from quantum channels using two-way communication protocols}
\author{Joonwoo Bae and Antonio Ac\'\i n}

\affiliation{ ICFO-Institut de Ciencies Fotoniques, Mediterranean
Technology Park, 08860 Castelldefels (Barcelona), Spain }
\date{\today}

\begin{abstract}
We provide a general formalism to characterize the cryptographic
properties of quantum channels in the realistic scenario where the
two honest parties employ prepare and measure protocols and the
known two-way communication reconciliation techniques. We obtain a
necessary and sufficient condition to distill a secret key using
this type of schemes for Pauli qubit channels and generalized
Pauli channels in higher dimension. Our results can be applied to
standard protocols such as BB84 or six-state, giving a critical
error rate of $20\%$ and $27.6\%$, respectively. We explore
several possibilities to enlarge these bounds, without any
improvement. These results suggest that there may exist weakly
entangling channels useless for key distribution using prepare and
measure schemes.
\end{abstract}

\maketitle

\section{Introduction}

Quantum Cryptography, that is, Quantum Key Distribution (QKD)
followed by one-time pad, is one of the most important quantum
information applications. The existing cryptographic methods using
classical resources base their security on technical assumptions
on the eavesdropper, often called Eve, capabilities, such as
finite computational power or bounded memory \cite{Maurer CKD}.
Contrary to all these schemes, the security proofs of QKD
protocols, e.g. the BB84 protocol \cite{BB84}, do not rely on any
assumption on Eve's power: they are simply based on the fact that
Eve's, as well as the honest parties' devices are governed by
quantum theory \cite{review}. Thus, well-established quantum
features, such as the monogamy of quantum correlations
(entanglement) or the impossibility of perfect cloning \cite{no
cloning}, make QKD secure. Actually, any possible quantum attack
by Eve would introduce errors and modify the expected quantum
correlations between the honest parties, Alice and Bob. The amount
of these errors can be estimated using public discussion, so the
honest parties can judge whether their quantum channel can be used
for secure QKD, or abort the insecure transmission otherwise.

The monogamy of entangled quantum states (see \cite{Terhal}) can
be simply illustrated in the scenario where two distant parties
know to share a two-qubit maximally entangled state, the so-called
\emph{ebit},
\begin{equation}\label{singlet}
    \ket{\Phi^+}=\frac{1}{\sqrt 2}(\ket{00}+\ket{11}) .
\end{equation}
Since the state is pure, it cannot be correlated with a third
eavesdropping party. So, Alice and Bob can safely map their ebit
into a \emph{secret bit} by just measuring in the computational
bases (see, Fig. \w{qcdist}). It is meant by secret bit a random
bit shared by Alice and Bob that is uncorrelated to Eve, namely
$P(A,B,E) = P(A,B)P(E)$ and $P(A=0,B=0)=P(A=1,B=1)= 1/2$, where
$P(A,B,E)$ denotes the probability distribution describing Alice,
Bob and Eve's correlations. Then, a simple QKD protocol could
consist of Alice locally preparing a state $\ket{\Phi^+}$, sending
half of this state through the channel to Bob, and then measuring
in the computational bases. However, any realistic channel between
Alice and Bob is in general noisy, so the state sent by Alice
interacts with the environment and is transformed into a mixed
state, $\rho_{AB}$. As a consequence of the noisy interaction with
the environment, Alice and Bob measurement outcomes are no longer
perfectly correlated. The honest parties then should know how to
deal with errors. They should safely assume that Eve has the power
to control all the environment, so all the errors are due to her
interaction with the sent states: the larger the observed error
rate, the larger Eve's information.

Entanglement distillation protocols \cite{entanglement
distillation} offer a possible solution to the problem of errors
or decoherence in the quantum channel. It is a technique that
allows two separate parties to transform by local operations and
classical communication (LOCC) many copies of a known entangled
mixed state into a fewer number of pure ebits. These ebits can
later be consumed to establish secret bits. However, entanglement
distillation protocols are by far not feasible with present-day
technology. This is because they require the use of a quantum
memory, a device able to store quantum states, and controlled coherent operations. Both techniques turn
out to be experimentally very challenging.

However, in order to establish secret bits, Alice and Bob do not
necessarily have to go through entanglement distillation. A much
more feasible family of protocols consist of the honest parties
measuring their quantum states at the single-copy level and then
applying classical distillation techniques to the obtained
measurement outcomes. We denote these \emph{SIngle-copy
Measurements plus ClAssical Processing} protocols as SIMCAP
\cite{equivalence in 2}.
Actually, it is well known that in the case of SIMCAP protocols,
the honest parties do not have to use entanglement at all for the
correlation distribution \cite{bbm}. Indeed, Alice's preparation
of the entangled two-qubit state plus measurement can be replaced
by the preparation of a one-qubit state that is sent trough the
noisy channel to Bob, who later measures it. That is, any SIMCAP
protocol in the entanglement picture is equivalent to a
\emph{prepare and measure} scheme \cite{bbm}, which is much more
feasible from an applied point of view. The BB84 and the six-state \cite{6 state} protocols constitute known examples of
\emph{prepare and measure} QKD schemes.

\begin{figure}
  \includegraphics[width=8cm]{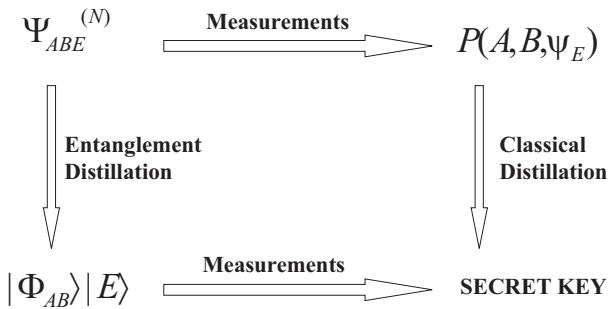}\\
  \caption{Schematic diagram for key distillation from
  quantum states: a secret key can be distilled either by
  entanglement distillation plus measurement, which is an experimentally challenging process, or
  by measurement plus classical processing of the outcomes, whose implementation is much more feasible.}
  \label{qcdist}
\end{figure}

Independently of the type of measurements or distillation
techniques employed in the protocol, a first and crucial step in
any QKD scheme consists of a \emph{tomographic process} by Alice
and Bob to obtain information about their connecting quantum
channel. By means of this process, Alice and Bob should conclude
whether the \emph{secrecy properties} of their channel are
sufficient to run a QKD protocol. In the standard formulation, the
cryptographic properties of quantum channels are referred to a
specific protocol. For instance, a standard problem is to
determine the critical quantum bit error rate (QBER) in the
channel such that key distillation is possible using one- or
two-way distillation techniques using the BB84 protocol. However,
it appears meaningful to identify and quantify the cryptographic
properties of a quantum channel by itself, independently of any
pre-determined QKD protocol. Indeed, this is closer to what
happens in reality, where the channel connecting Alice and Bob is
fixed. Therefore, after the tomographic process, the two honest
parties should design the protocol which is better tailored to the
estimated channel parameters. In this sense, it is well known that
no secure QKD can be established using entanglement-breaking
channel \cite{GW,CLL}, while the detection of entanglement already
guarantees the presence of some form of secrecy \cite{secret
correlation}. Beyond these two results, little is known about
which channel properties are necessary and/or sufficient for
secure QKD.

In the present work, we analyze the cryptographic properties of
quantum channels when Alice and Bob employ QKD schemes where (i)
the correlation distribution is done using prepare and measure
techniques and (ii) the key distillation process uses the standard
one-way and two-way classical protocols. Indeed, these are the
techniques presently used in any realistic QKD implementation. It
should be clear, then, that none of the protocols considered here
require the use of entangled particles. However, for the sake of
simplicity, we perform our analysis in the completely equivalent
entanglement picture. As it becomes clearer below, the problem
then consists of identifying those quantum states that can be
distilled into secret bits by SIMCAP protocols restricted to the
known distillation techniques. A first step in this direction has
recently been given in \cite{collective security}. There, a rather
easily computable and powerful necessary condition for secure QKD
is derived, which is shown to be sufficient against the so-called
collective attacks (see below). In general, the derived necessary
condition is more restrictive than the entanglement condition. In
this work, we first rederive the security condition of
\cite{collective security}, improving the security analysis. Since
collective attacks have been proven to be as powerful as general
attacks \cite{Renner}, our condition actually applies to any
attack. We show how to apply this condition to the standard BB84
and six-state protocols. Next, we explore several possibilities to
improve the obtained security bounds. Remarkably, all these
alternatives fail, which suggests the existence of non-distillable
entangled states under general SIMCAP protocols. Then, we move to
higher dimensional systems, also called \emph{qudits}, and extend
the results to generalized Bell diagonal qudit channels. The
obtained security condition turns out to be tight for the
so-called $(d+1)$- and 2-bases protocol of Ref. \cite{cbkg}.

The article is organized as follows. Section \ref{secprot0}
defines what we call \emph{realistic} protocols. In section
\ref{secattacks}, we introduce and classify several eavesdropping
attacks. Exploiting the connection between QKD and the de Finetti
theorem established by Renner \cite{Renner}, we can
restrict the security analysis to the so-called collective
attacks, where Eve applies the same interaction to each quantum
state. Then, we briefly review some of the existing security
bounds for the two most commonly used prepare and measure
protocols, BB84 and six-state (section \ref{revbounds}). In the
next section, we derive the announced security condition for qubit
channels and apply it to the two mentioned protocols. We then show
that neither pre-processing nor coherent quantum operations by one
of the parties improves the obtained security bounds. In section
\ref{secqudits}, we move to higher dimensional systems, extending
the security conditions to generalized Bell diagonal channels.
Then, we apply this condition to the $(d+1)$- and 2-bases protocols
of \cite{cbkg}, which can be understood as the natural
generalization to qudits of the BB84 and the six-state protocols,
and prove the tightness for these protocols. Finally, section
\ref{secconcl} summarizes the main results and open questions
discussed in this work. Most of the technical details are left for
the appendices.

\section{Realistic Protocol}
\label{secprot0}

There exist plenty of QKD protocols in the literature. Here, we
restrict our considerations to what we call realistic protocols
where Alice prepares and sends states from a chosen basis to Bob,
who measures in another (possibly different) basis. This
establishes some classical correlations between the two honest
parties. Of course this process alone is clearly insecure, since
Eve could apply an intercept resend strategy in the same basis as
Alice's state preparation, acquiring the whole information without
being detected. Therefore, from time to time, Alice and Bob should
change their state preparation and measurements to monitor the
channel and exclude this possibility. Alice and Bob announce these
symbols to extract information about their channel, so these
instances do not contribute to the final key rate. Indeed these
symbols are waisted in the tomographic process previously
mentioned. However, in the limit of large sequences, the fraction
of cases where Alice and Bob monitor the channel can be made
negligible in comparison with the key length, but still sufficient
to have a faithful description of some channel parameters, such as
the QBER \cite{note}. The states sent by Alice will be transformed
into a mixed state because of Eve's interaction. This decoherence
will produce errors in the measurement values obtained by Bob. The
security analysis aims at answering whether the observed
decoherence in the channel is small enough to allow Alice and Bob
distilling a secret key. We call these protocols realistic in the
sense that they do not involve experimentally difficult quantum
operations, such as coherent measurements, quantum memories or the
generation of entangled particles. The establishment of
correlations is done by just generating one-qubit states and
measuring them in two or more bases. Additionally, one could think
of including a filtering single-copy measurement on Bob's side.
This operation is harder than a standard projective measurement,
but still feasible with present-day technology \cite{povmexp}.

The above scenario can be explained in the completely equivalent
entanglement-based scenario \cite{bbm}, that turns out to be much
more convenient for the theoretical analysis. In the
entanglement-based scheme, the information encoding by Alice is
replaced by generating and measuring half of a maximally entangled
state. That is, Alice first locally generates a maximally
entangled two-qubit state and sends half of it to Bob through the
channel. A mixed state $\rho_{AB}$ is then shared by the two
honest parties, due to the interaction with the environment
controlled by Eve. Now, Alice and Bob measure in two bases to map
their quantum correlations into classical correlations. For
instance, if Alice and Bob measure in the computational bases, the
QBER simply reads \bea \epsilon_{AB} = \langle 01
|\rho_{AB}|01\rangle + \langle 10 |\rho_{AB}| 10 \rangle .
\nonumber \eea It can be imposed that Alice's
local state cannot be modified by Eve, since the corresponding
particle never leaves Alice's lab, which is assumed to be secure.
It has to be clear that the techniques of \cite{bbm} imply the
equivalence between SIMCAP protocols on entangled states and
prepare and measure QKD schemes: the correlation distribution is,
from the secrecy point of view, identical. This equivalence, for
instance, is lost if one considers entanglement distillation
protocols for QKD, where the particles are measured by the honest
parties after applying coherent quantum operations.

\begin{figure}
  \includegraphics[width=8cm]{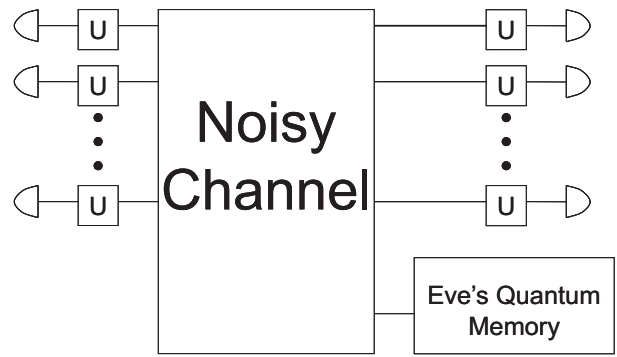}\\
  \caption{A tripartite pure state is prepared by Eve, who send two of the particles to Alice and Bob and keeps one. From Alice and Bob viewpoint the situation resembles a standard noisy channel. The honest parties perform measurements at the single copy level, possibly with some preliminary filtering step. Eve keeps her quantum states and can arbitrarily delay her collective measurement.}\label{scheme}
\end{figure}

\subsection{Classical key distillation}

After the correlation distribution, either using prepare and
measure or SIMCAP protocols, Alice and Bob share partially secret
correlations to be distilled into the perfect key. The problem of
distilling noisy and partially secret correlations into a secret
key has not been completely solved. Recently, general lower bounds
to the distillable secret-key rate by means of error correction
and privacy amplification using one-way communication have been
obtained in \cite{DV}. In case the correlations are too noisy for
the direct use of one-way distillation techniques, Alice and Bob
can before apply a protocol using two-way communication. The
obtained correlations after this two-way process may become
distillable using one-way protocols. Much less is known about key
distillation using two-way communication. Here we mostly apply the
standard two-way communication protocol introduced by Maurer in
\cite{Maurer CAD}, also known as classical advantage distillation
(CAD). Actually, we analyze the following two slightly different
CAD protocols:

\begin{itemize}
    \item \emph{CAD1.} Alice and Bob share a list of correlated bits.
    Alice selects $N$ of her bits that have the same value and
    publicly announces the position of these symbols.
    Bob checks whether his corresponding symbols are also equal. If
    this is the case, Bob announces to Alice that he accepts,
    so they use the measurement values (they are all the same) as a
    bit for the new list. Otherwise, they reject the $N$ values and
    start again the process with another block.

    \item \emph{CAD2.} Alice locally generates a random bit $s$.
    She takes a block of $N$ of her bits, $A$, and computes the vector
    \be X =(X_{1},\cdots,X_{N}) \label{ab} \ee such that $A_i +
    X_i= s$. She then announces the new block $X$ through the public
    and authenticated classical channel. After receiving $X$, Bob adds it
    to his corresponding block, $B + X$, and accepts whenever all the
    resulting values are the same. If not, the symbols are
    discarded and the process is started again, as above.
\end{itemize}

These protocols are equivalent in classical cryptography and in
the completely general quantum scenario. Nevertheless, it is shown
in section \ref{secineq} that they are different in some
particular, but still relevant, scenarios. In what follows, we restrict the analysis to key distillation protocols consisting of CAD followed by standard one-way error correction and privacy amplification. Thus, it is important to keep in mind that any security claim is referred to this type of key-distillation protocols. Although these are the protocols commonly used when considering two-way reconciliation techniques, their optimality, at least in terms of robustness, has not been proven.

\section{Eavesdropping Strategies}
\label{secattacks}

After describing Alice and Bob's operations, it is now time to
consider Eve's attacks. With full generality, we suppose that Eve
has the power to control all the environment. That is, all the
information that leaks out through the channel connecting Alice
and Bob goes to Eve, so all the decoherence seen by Alice and Bob
is introduced by her interaction. Following Ref. \cite{collective
security}, eavesdropping strategies can be classified into three
types: (i)~individual, (ii)~collective and (iii)~coherent. Once
more, although most of the following discussion is presented in
the entanglement picture, the same conclusions apply to the
corresponding prepare and measure scheme.

\subsection{Individual attacks}

In an individual attack Eve is assumed to apply the same
interaction to each state, without introducing correlations among
copies, and measure her state right after this interaction. In
this type of attacks, all three parties immediately measure their
states, since no one is supposed to have the ability to store
quantum states. Therefore, they end up sharing
classical-classical-classical (CCC) correlated measurement
outcomes \cite{notation CCQ}, described by a probability
distribution $P(A,B,E)$. In this case, standard results from
Classical Information Theory can be directly applied. For
instance, it is well known that the secret-key rate using one-way
communication, $K_\rightarrow$, is bounded by so-called
Csisz\'ar-K\"{o}rner bound \cite{CK theorem},
\begin{equation}\label{CKbound}
    K_\rightarrow\geq I(A:B)-I(A:E)  .
\end{equation}
Here $I(A:B)$ denotes the classical mutual information between the
measurement outcomes $A$ and $B$. It reads
\begin{equation}
    I(A:B)=H(A)-H(A|B) ,
\end{equation}
where $H$ denotes the standard Shannon entropy. In this type of
attacks, Eve's interaction can be seen as a sort of asymmetric
cloning \cite{asyclon} producing two different approximate copies,
one for Bob and one for her. This cloning transformation reads
$U_{BE}:|\Phi^+ \rangle_{AB} |E \rangle \to |\Psi \rangle_{ABE} $
where $\rho_{AB} = \tr_{E} |\Psi\rangle\langle \Psi|_{ABE} $. It
has been shown that in the case of two qubits, two honest parties
can distill a secret key secure against any individual attacks
whenever their quantum state $\rho_{AB}$ is entangled
\cite{equivalence in 2}.

It is clear that to prove security against individual attacks is
not satisfactory from a purely theoretical point of view. However,
we believe it is a relevant issue when dealing with realistic
eavesdroppers. Assume Eve's quantum memory decoherence rate is
nonzero and the honest parties are able to estimate it. Then, they
can introduce a delay between the state distribution and the
distillation process long enough to prevent Eve keeping her states
without errors. Eve is then forced to measure her states before
the reconciliation, as for an individual attack.

\subsection{Collective Attacks}

Collective attacks represent, in principle, an intermediate step
between individual and the most general attack. Eve is again
assumed to apply the same interaction to each quantum state, but
she has a quantum memory. In other words, she is not forced to
measure her state after the interaction and can arbitrarily delay
her measurement. In particular, she can wait until the end of the
reconciliation process and adapt her measurement to the public
information exchanged by Alice and Bob. After a collective attack,
the two honest parties share $N$ independent copies of the same
state, $\rho_{AB}^{\otimes N}$, where no correlation exists from
copy to copy. Without losing generality, the full state of the
three parties can be taken equal to $|\psi\rangle_{ABE}^{\otimes
N}$, where \bea |\psi\rangle_{ABE}=(I_{A}\otimes
U_{BE})|\Phi^{+}\rangle_{AB}|E\rangle .\eea After a collective
attack, and the measurements by Alice and Bob, the three parties
share classical-classical-quantum (CCQ) correlations, described by a
state
\begin{equation}
\label{collst}
    \sum_{a,b} [a]\otimes[b]\otimes[e_{ab}] ,
\end{equation}
where $a$ and $b$ denote Alice and Bob's measurement outcomes
associated to the measurement projectors $[a]$ and $[b]$.
Throughout this paper, square brackets denote one-dimensional
projector, e.g. $[\psi] = |\psi\rangle \langle\psi|$. Note that
$[e_{ab}]$ is not normalized, since
$\ket{e_{ab}}=\bra{ab}\psi\rangle_{ABE}$ and
$p\,(a,b)=\tr[e_{ab}]$.

The following result, obtained in \cite{DV,RNB}, is largely used
in the next sections. After a collective attack described by a
state like (\ref{collst}), Alice and Bob's one-way distillable key
rate satisfies \bea K_{\rightarrow} \geq I(A:B) - I(A:E) .
\label{dw} \eea Here, the correlations between Alice and Bob's
classical variables are again quantified by the standard mutual
information, $I(A:B)$. The correlations between Alice's classical
and Eve's quantum variables, $A$ and $E$, are quantified by the
Holevo quantity,
\begin{equation}\label{holevo}
    I(A:E) = S(E)-S(E|A) ,
\end{equation}
where $S$ denotes the Shannon entropy, so $S(E)=S(\rho_E)$ and
$S(E|A)=\sum_a p\,(a) S(\rho_E|A=a)$. Actually the ``same"
equation (\ref{dw}) applies when Bob is also able to store quantum
states and the three parties share classical-quantum-quantum (CQQ)
correlations. In this case, both mutual information quantities
between Alice's classical variable, $A$, and Bob's and Eve's
quantum states, denoted by $B$ and $E$, should be understood as
Holevo quantities \cite{DV}. Notice the similarities between
(\ref{CKbound}) and (\ref{dw}). Indeed, the obtained bounds
represent a natural generalization of the CK-bound to the CCQ and
CQQ correlations scenarios.

\subsection{General Attacks and the de Finetti Theorem}

Finally, one has to consider the most general attack where Eve can
perform any kind of interaction. In this case, Alice and Bob
cannot assume to share $N$ copies of the same quantum state.
Compared to the previous attacks, there did not exist nice bounds
for the extractable key-rate under general attacks. However, very
recently a dramatic simplification on the security analysis of QKD
protocols under general attacks has been achieved by means of the
so-called de Finetti theorem \cite{Renner}. Indeed, Renner has
proven that general attacks cannot be more powerful than
collective attacks in any protocol that is symmetric in the use of
the quantum channel. This provides a huge simplification in
security proofs, since by means of the de Finetti arguments (see
\cite{Renner} for more details), Alice and Bob can safely assume
to share $N$ copies of a quantum state consistent with their
tomographic process, and then apply the existing bounds for this
scenario. Note that the de Finetti theorem should also be employed
if one wants to use entanglement distillation as a key
distillation technique. In what follows, then, we can restrict our
analysis to collective attacks, without underestimating Eve's
capabilities.

\subsection{Review of the existing Security Bounds}
\label{revbounds}

Finally, we would like to summarize the existing security bounds
for the two most known QKD protocols, BB84 and six-state. These
bounds are usually stated in terms of the critical QBER such that
key distillation is possible. Of course, these bounds depend on
the type of key distillation techniques employed by the honest
parties. Since the first general security proof of BB84 by Mayers
\cite{Mayers}, security bounds have been constantly improved.
Using a quantum error-correction (of bit-flip and phase-inversion)
description of classical one-way error-correction and privacy
amplification, Shor and Preskill showed the general security of
BB84 whenever $QBER<11\%$ \cite{Shor and Preskill}. Later, Lo
adapted their proof to 6-state protocol obtaining a critical QBER
of $12.7\%$ \cite{hklo}. More recently, Kraus, Renner, and Gisin
have improved these values by introducing some classical
pre-processing by the two honest parties, obtaining critical
QBER's of $12.4\%$ for the BB84 and $14.1\%$ for the six-state
protocol \cite{RNB}. More recently, the bound for BB84 has been
improved up to $12.9\%$ in Ref. \cite{SRS}. On the other hand, the
known upper bounds on the critical QBER are slightly higher than these lower
bounds, so the exact value for the critical QBER remains as an
open question.

\begin{figure}
  \includegraphics[width=8cm]{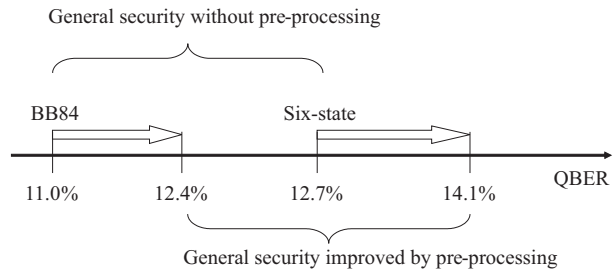}\\
  \caption{Security bounds for QKD protocols using key distillation techniques with one-way communication:
  based on the analogy between these techniques and quantum error correction,
  the security bounds for the BB84 and the six-state protocols are $11\%$ and $12.7\%$ respectively. These bounds have later been improved by information-theoretic
  considerations up to $12.4\%$ and $14.1\%$. The improvement is achieved using some classical
  pre-processing by one of the parties.}\label{one-way}
\end{figure}

The honest parties however can apply CAD to their outcomes before
using one-way key-distillation techniques and improve these
bounds. The whole process can now be mapped into a two-way
entanglement distillation protocol. Based on this analogy,
Gottesman and Lo have obtained that secure QKD is possible
whenever the QBER is smaller than $18.9\%$ and $26.4\%$ for the
BB84 and six-state protocol, respectively \cite{go-lo}. Chau has
improved these bounds up to $20.0\%$ and $27.6\%$ respectively
\cite{chau}. The generalization of the formalism \cite{RNB} to two-way communication has also been done by Kraus, Branciard and Renner \cite{BKR}. We show in the next sections (see also
\cite{collective security}) that, for larger QBER, no protocol
consisting of CAD followed by one-way distillation techniques
works. So, contrary to what happens in the case of one-way
communication, there is no gap between the lower and upper bounds
for secure key distribution using the BB84 and six-state schemes, under the considered reconciliation techniques.

\begin{figure}
  \includegraphics[width=8cm]{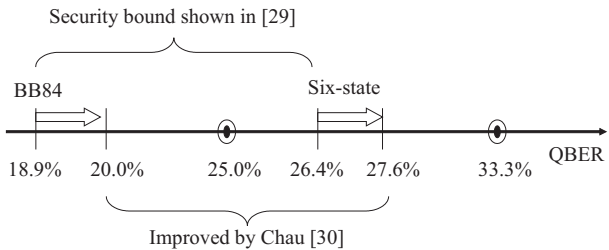}\\
  \caption{Security bounds for QKD protocols using two-way followed by one-way communication techniques:
  based on the analogy between the two-way plus one-way communication and two-way entanglement distillation protocol,
  general security bounds of the BB84 and the six-state protocols are given by $18.9\%$ and $26.4\%$ respectively \cite{go-lo}. Later, Chau improved the error correction method and the
  bounds are moved to $20.0\%$ and $27.6\%$ \cite{chau}. In sections IV and V, we show that those bounds are tight.
  Note that the key distillability condition is stronger than the entanglement
  condition, which is $25.0 \%$ and $33.3 \%$ for the BB84 and the
  six-state protocols.}\label{two-way}
\end{figure}

\section{Secrecy Properties of Qubit Channels}
\label{secprop}

After reviewing the main ideas and previous results used in what
follows, we are in position of deriving our results. Consider the
situation where Alice and Bob are connected by a qubit channel.
Alice locally prepares a maximally entangled state of two qubits
and sends half of it through the channel. Then, both parties
measure the state. By repetition of this process, they can obtain
a complete, or partial, characterization of their channel, up to
some precision. Indeed, there exists a correspondence between a
channel, $\Upsilon$, and the state
\begin{equation}\label{stchannel}
    (\one\otimes\Upsilon)\ket{\Phi^+}=\rho_{AB} .
\end{equation}
Now, the parties agree on a pair of bases, that will later be used
for the raw key distribution. They repeat the same process but now
measure almost always in these bases. However, with small
probability, they have to change their measurement to the previous
tomographic process in order to check the channel. After public
communication, they discard the asymptotically negligible fraction
of symbols where any of them did not use the right basis and
proceed with the key distillation. In what follows, we provide a
security analysis of this type of schemes. Two important points
should be mentioned again: (i) as said, these schemes can be
easily transformed into a prepare and measure protocol, without
entanglement and (ii) using de Finetti theorem, Alice and Bob can
restrict Eve to collective attacks. In other words, they can
assume to share $N$ independent copies of the same state,
$\rho_{AB}^{\otimes N}$, that is, the channel does not introduce
correlation between the states. The goal, then, consists of
finding the optimal SIMCAP protocol for the state $\rho_{AB}$, or
equivalently, the best prepare and measure scheme for the channel
$\Upsilon$.

Generically, $\rho_{AB}$ can be any two-qubit state. However, no
key distillation is possible from separable states, so Alice and
Bob abort their protocol if their measured data are consistent
with a separable state \cite{secret correlation}. We can assume,
if the state preparation is done by Alice, that her local state,
$\rho_A$, cannot be modified by Eve. In our type of schemes, this
state is equal to the identity. Although our techniques can be
used in the general situation, we mostly restrict our analysis to
the case where Bob's state is also equal to the identity. This is
likely to be the case in any realistic situation, where the
channel affects with some symmetry the flying qubits. This
symmetry is reflected by the local state on reception, i.e.
$\rho_B=\one$. In the qubit case, the fact that the two local
states are completely random simply means that the global state
$\rho_{AB}$ is Bell diagonal,
\begin{equation}\label{bdstate}
    \rho_{AB}=\lambda_1[\Phi_1]+\lambda_2[\Phi_2]+
    \lambda_3[\Phi_3]+\lambda_4[\Phi_4] ,
\end{equation}
where $\sum_{j}\lambda_{j} = 1$, $\lambda_{j} > 0$, and
\begin{eqnarray}
  \ket{\Phi_1} &=& \frac{1}{\sqrt 2}(\ket{00}+\ket{11})\nonumber \\
  \ket{\Phi_2} &=& \frac{1}{\sqrt 2}(\ket{00}-\ket{11})\nonumber \\
  \ket{\Phi_3} &=& \frac{1}{\sqrt 2}(\ket{01}+\ket{10})\nonumber \\
  \ket{\Phi_4} &=& \frac{1}{\sqrt 2}(\ket{01}-\ket{10})
\end{eqnarray}
define the so-called Bell basis. Or in other words, $\Upsilon$ is
a Pauli channel. Pauli channels are very useful, as it will become
clearer below, in the analysis of the BB84 and six-state
protocols.

It is also worth mentioning here that Alice and Bob can always
transform their generic state $\rho_{AB}$ into a Bell diagonal
state by single-copy filtering operations. Actually, this
operation is optimal in terms of entanglement concentration.
Indeed, it maximizes the entanglement of formation of any state
$\rho_{AB}'\propto(F_{A}\otimes F_{B})\rho (F_{A}^{\dagger}\otimes
F_{B}^{\dagger})$ obtained after LOCC operations of a single copy
of $\rho_{AB}$ \cite{filtering maximize}. This filtering operation
succeeds with probability $\tr(F_{A}\otimes F_{B})\rho
(F_{A}^{\dagger}\otimes F_{B}^{\dagger})$. If $\rho_{AB}$ is
already in a Bell-diagonal form, it remains invariant under the
filtering operation. Alternatively, Alice and Bob can also map
their state into a Bell diagonal state by a depolarization
protocol, where they apply randomly correlated change of basis,
but some entanglement may be lost in this process. In view of all
these facts, in what follows we mainly consider Bell diagonal
states.

It is possible to identify a canonical form for these states. This
follows from the fact that Alice and Bob can apply local unitary
transformation such that
\bea
    \lambda_1= \max_i \lambda_i ,\qquad
    \lambda_2= \min_i \lambda_i \ . \label{minmax}
\eea Indeed, they can permute the Bell basis elements by
performing the following operations
\bea T([\Phi_{1}
]\leftrightarrow [\Phi_{2}])&=&2^{-1}i(\openone-i\sigma_z)
\otimes(\openone-i\sigma_z),\nonumber \\
T([\Phi_{2}]\leftrightarrow[
\Phi_{3}])&=&2^{-1}(\sigma_x+\sigma_z)
\otimes(\sigma_x+\sigma_z),\nonumber \\
T([\Phi_{3}] \leftrightarrow
[\Phi_{4}])&=&2^{-1}(\openone+i\sigma_z)
\otimes(\openone-i\sigma_z). \label{lu} \eea Once the state has
been casted in this canonical form, Alice and Bob measure it in
the computational basis. The choice of the computational bases by
Alice and Bob will be justified by our analysis. Indeed, once a
Bell-diagonal state has been written in the previous canonical
form, the choice of the computational bases seems to maximize the
secret correlations between Alice and Bob, although, in general,
they may not maximize the total correlations.

Before Alice and Bob' measurements, the global state including Eve
is a pure state that purifies Alice and Bob's Bell diagonal state,
that is, \bea |\Psi \rangle_{ABE} =
\sum_{j=1}^{4}\sqrt{\lambda_{j}}|\Phi_{j}\rangle |j\rangle_{E}
\label{ABE} \eea where $|j\rangle_{E}$ define an orthonormal basis
on Eve's space. All the purifications of Alice-Bob state are
equivalent from Eve's point of view, since they only differ from a
unitary operation in her space. After the measurements, Alice, Bob
and Eve share CCQ correlations. In the next sections we study when
these correlations can be distilled into a secure key using the
standard CAD followed by one-way distillation protocols. We first
obtain a sufficient condition for securtiy, using the lower bounds
on the secret-key rate given above, c.f. \w{dw}. Then, we compute
a necessary condition that follows from a specific eavesdropping
attack. It is then shown that the two conditions coincide, so the
resulting security condition is necessary and sufficient, under
the mentioned distillation techniques. Next, we apply this
condition to two known examples, the BB84 and the six-state
protocols. We finally discuss several ways of improving the
derived condition, by changing the distillation techniques,
including classical pre-processing by the parties or one-party's
coherent quantum operations.

\subsection{Sufficient condition}
\label{secsuffcond}

In this section we will derive the announced sufficient condition
for security using the lower bound on the secret-key rate of Eq.
\w{dw}. Just before the measurements, the honest parties share a
Bell diagonal state (\ref{bdstate}). This state is entangled if
and only if $\sum_{j=2}^{4}\lambda_{j}<\lambda_{1}$, which follows
from the fact that the positivity of the partial transposition is
a necessary and sufficient condition for separability in $2\times
2$ systems \cite{ppt}. When Alice and Bob measure in their
computational bases, they are left with classical data
$[i,j~]_{AB}$($i,j\in \{0,1\}$) whereas Eve still holds a quantum
correlated system $|e_{i,j}\rangle_{E}$. The CCQ correlations they
share are described by the state (up to normalization)
  \be
  \label{ccqcorr}
  \rho_{ABE} \varpropto \sum_{i,j}\ [i,j~]_{AB}\otimes[\widetilde{e_{i,j}}]_{E},
  \ee
where Eve's states are
  \bea
  \ket{\widetilde{e_{0,0}}} & = & \sqrt{\lambda_1}\ket{1} + \sqrt{\lambda_2}\ket{2}
  \nonumber\\
  \ket{\widetilde{e_{0,1}}} & = & \sqrt{\lambda_3}\ket{3} + \sqrt{\lambda_4}\ket{4}
  \nonumber\\
  \ket{\widetilde{e_{1,0}}} & = & \sqrt{\lambda_3}\ket{3} - \sqrt{\lambda_4}\ket{4}
  \nonumber\\
  \ket{\widetilde{e_{1,1}}} & = & \sqrt{\lambda_1}\ket{1} -
  \sqrt{\lambda_2}\ket{2} ,
  \eea
and the corresponding states without tilde denote the normalized
vectors. So, after the measurements, Alice and Bob map
$\rho_{AB}^{\otimes N}$, into a list of measurement outcomes,
whose correlations are given by $P_{AB}(i,j)$, where
\begin{equation}
    P_{AB}(i,j)=\bra{ij}\rho_{AB}\ket{ij} .
\end{equation}
This probability distribution reads as follows:
\begin{center}
\begin{tabular}{|c|c|c|}
  \hline
  A $\setminus$ B & 0 & 1 \\
  \hline
  0 & $(1-\epsilon_{AB})/2$ & $\epsilon_{AB}/2$ \\
\hline
  1 & $\epsilon_{AB}/2$ & $(1-\epsilon_{AB})/2$ \\
  \hline
\end{tabular}
\end{center}
Here, $\epsilon_{AB}$ denotes the QBER, that is,
\be \epsilon_{AB}
= \langle 01 |\rho_{AB}|01 \rangle +\langle 10
|\rho_{AB}|10 \rangle  \\
 =  \lambda_{3} + \lambda_{4} . \\
\ee

Alice and Bob now apply CAD to a block of $N$ symbols. Eve listens
to the public communication that the two honest parties exchange.
In particular, she has the position of the $N$ symbols used by
Alice in \w{ab}, in case the honest parties use $CAD1$ or the
$N$-bit string $X$ for $CAD2$. In the second case, Eve applies to
each of her symbols the unitary transformation
  \be
  U_{i}=[1]_E+(-1)^{X_{i}}[2]_E+[3]_E+(-1)^{X_{i}}[4]_E.
  \label{transform}
  \ee
This unitary operation transforms $|e_{i,j}\rangle_{E}$ into
$|e_{s,j}\rangle_{E}$ where $s$ is the secret bit generated by
Alice. If Alice and Bob apply CAD1, Eve does nothing. In both
cases, the resulting state is \bea \rho_{ABE}^{N} & = &
\frac{(1-\epsilon_{N})}{2} \sum_{s=0,1}[s,s]_{AB}
\otimes [e_{s,s}]^{\otimes N} + \nonumber \\
&& \frac{\epsilon_{N}}{2} \sum_{s=0,1}[s,s+1]_{AB} \otimes
[e_{s,s+1 }]^{\otimes N} ,\label{tripartite} \eea where
$\epsilon_{N}$ is Alice-Bob error probability after CAD, \bea
\epsilon_{N} =\frac{\epsilon_{AB}^{N}}
{\epsilon_{AB}^{N}+(1-\epsilon_{AB})^{N}}\leq\left(\frac{\epsilon_{AB}}
{1-\epsilon_{AB}}\right)^{N} , \label{epN} \eea and the last
inequality tends to an equality when $N\to\infty$. That is,
whatever the advantage distillation protocol is, i.e. either CAD1
or CAD2, all the correlations among the three parties before the
one-way key extraction step are described by the state
\w{tripartite}.

We can now apply Eq. (\ref{dw}) to this CQQ state. The probability
distribution between Alice and Bob has changed to
\begin{center}
\begin{tabular}{|c|c|c|}
  \hline
  A $\setminus$ B & 0 & 1 \\
  \hline
  0 & $(1-\epsilon_{N})/2$ & $\epsilon_{N}/2$ \\
\hline
  1 & $\epsilon_{N}/2$ & $(1-\epsilon_{N})/2$ \\
  \hline
\end{tabular}
\end{center}
where it can be seen that Alice and Bob have improved their
correlation. The CAD protocol has changed the initial probability
distribution $P(A,B)$, with error rate $\epsilon_{AB}$, into
$P^{'}(A,B)$, with error rate $\epsilon_{N}$. The mutual
information between Alice and Bob $I(A:B)$ is easily computed from
the above table. $I(A:E)$ can be derived from (\ref{tripartite}),
so, after some algebra, the following equality is obtained
  \bea
 && I(A:B)-I(A:E)= 1 -h(\epsilon_{N}) \nonumber\\
 &&- (1-\epsilon_{N})\, h\!\left( \frac{1-\Lambda_{\rm eq}^M}{2} \right)
  -\epsilon_{N}    \, h\!\left( \frac{1-\Lambda_{\rm dif}^M}{2} \right) ,
  \label{calcul}
  \eea
where
  \bea
 \displaystyle \Lambda_{\rm eq} &=&
 \frac{\lambda_1-\lambda_2}{\lambda_1+\lambda_2}=
 |\langle e_{0,0}|e_{1,1}\rangle| \nonumber\\
 \Lambda_{\rm dif} &=&
 \frac{|\lambda_3-\lambda_4|}{\lambda_3+\lambda_4}=
 |\langle e_{1,0}|e_{0,1}\rangle|,\label{errbob}
  \eea
$h(x)=-x \log_2 x - (1-x)\log_2 (1-x)$ is the binary entropy, and
the subscript `eq' (`dif') refers to the resulting value of Alice
being equal to (different from) that of Bob.

Let's compute this quantity in the limit of a large number of
copies, $N\gg 1$, where $\epsilon_N,\Lambda_{\rm eq},\Lambda_{\rm
dif}\ll 1$. It can be seen that in this limit
\begin{eqnarray}
\label{largeN}
  I(A:B) &\approx& 1+\epsilon_N\log \epsilon_N \nonumber\\
  I(A:E) &\approx& 1-\frac{1}{\ln 4}\Lambda_{\rm eq}^{2N} .
\end{eqnarray}
The security condition follows from having positive value of the
Eq. \w{calcul}, which holds if \bea |\langle e_{0,0}|e_{1,1}
\rangle|^{2} > \frac{\epsilon_{B}}{1-\epsilon_{B}}
.\label{security}\eea More precisely, if this condition is
satisfied, Alice and Bob can always establish a large but finite
$N$ such that Eq. (\ref{calcul}) becomes positive. Eq.
(\ref{security}) can be rewritten as
\begin{equation}\label{security2}
    (\lambda_1+\lambda_2)(\lambda_3+\lambda_4)<(\lambda_1-\lambda_2)^2
    .
\end{equation}
Therefore, whenever the state of Alice and Bob satisfies the
security condition \w{security} above, they can extract from
$\rho_{AB}$ a secret key with our SIMCAP protocol. This gives the
searched sufficient condition for security for two two-qubit Bell
diagonal states or, equivalently, Pauli channels. Later, it is
proven that whenever condition \w{security} does not hold, there
exists an attack by Eve such that no standard key-distillation
protocol works.

Condition (\ref{security}) has a clear physical meaning. The r.h.s
of (\ref{calcul}) quantifies how fast Alice and Bob's error
probability goes to zero when $N$ increases. In the same limit,
and since there are almost no errors in the symbols filtered by
the CAD process, Eve has to distinguish between $N$ copies of
$\ket{e_{0,0}}$ and $\ket{e_{1,1}}$. The trace distance between
these two states provides a measure of this distinguishability. It
is easy to see that for large $N$
\begin{eqnarray}
    \tr|[e_{0,0}]^{\otimes N}-[e_{1,1}]^{\otimes
    N}|&=&2\sqrt{1-|\bra{e_{0,0}}e_{1,1}\rangle|^{2N}}\nonumber\\
    &\approx& 2-|\bra{e_{0,0}}e_{1,1}\rangle|^{2N} .
\end{eqnarray}
Thus, the l.h.s. of (\ref{calcul}) quantifies how the
distinguishability of the two quantum states on Eve's side after
CAD increases with $N$. This intuitive idea is indeed behind the
attack described in the next section.



Once this sufficient condition has been obtained, we can justify
the choice of the computational bases for the measurements by
Alice and Bob when sharing a state (\ref{bdstate}). Note that the
same reasoning as above can be applied to any choice of bases. The
derived security condition simply quantifies how Alice-Bob error
probability goes to zero with $N$ compared to Eve's
distinguishability of the $N$ copies of the states $\ket{e_{0,0}}$
and $\ket{e_{1,1}}$, corresponding to the cases $a=b=0$ and
$a=b=1$. The obtained conditions are not as simple as for
measurements in the computational bases, but they can be easily
computed using numerical means. One can, then, perform a numerical
optimization over all choice of bases by Alice and Bob. An
exhaustive search shows that computational bases are optimal for
this type of security condition. It is interesting to mention that
the bases that maximize the classical correlations, or minimize
the error probability, between Alice and Bob do not correspond to
the computational bases for all Bell diagonal states
(\ref{bdstate}). Thus, these bases optimize the {\sl secret
correlations} between the two honest parties, according to our
security condition, although they may be not optimal for classical
correlations.

\begin{figure}
  \includegraphics[width=8cm]{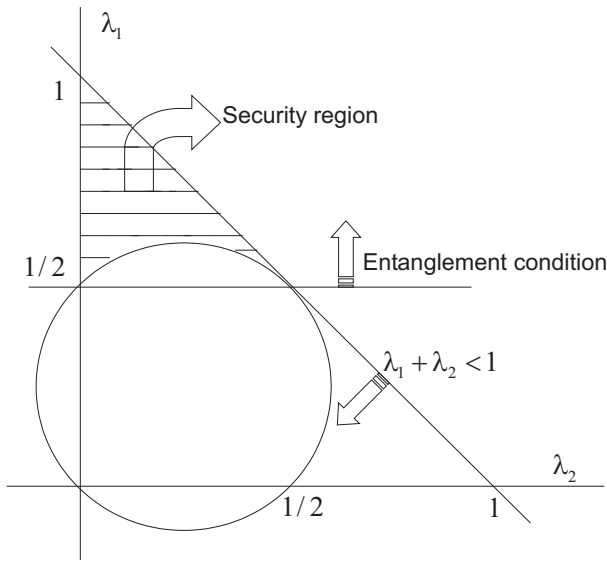}\\
  \caption{Graphical depiction of the security condition \w{security2}: the security region
  is defined by the intersection of the entanglement condition $\lambda_{1}>1/2$, the normalization condition $\lambda_{1}+\lambda_{2}<1$,
  and the security condition \w{security2}. 
}\label{se}
\end{figure}


\subsection{Necessary condition}
\label{sectatt}

After presenting the security condition (\ref{security}), we now
give an eavesdropping attack that breaks our SIMCAP protocol
whenever this condition does not hold. This attack is very similar
to that in Ref. \cite{sing}.

Without loss of generality, we assume that all the communication
in the one-way reconciliation part of the protocol goes from Alice
to Bob. In this attack, Eve delays her measurement until Alice and
Bob complete the CAD part of the distillation protocol. Then, she
applies on each of her systems the two-outcome measurement defined
by the projectors
  \be
  F_{\rm eq}=[1]_E+[2]_E ,\quad
  F_{\rm dif}=[3]_E+[4]_E .
  \label{projectors2}
  \ee
According to \w{tripartite}, all $N$ measurements give the same
outcome. If Eve obtains the outcome corresponding to $F_{\rm eq}$,
the tripartite state becomes (up to normalization)
 \be
 [00]_{AB}\otimes [e_{0,0}]_E^{\otimes N}
 +[11]_{AB}\otimes [e_{1,1}]_E^{\otimes N} .
  \label{AeqB}
 \ee

In order to learn $s_A$, Alice's bit, she has to discriminate
between the two pure states $|e_{0,0}\rangle ^{\otimes N}$ and
$|e_{1,1}\rangle^{\otimes N}$. The minimum error probability in
such discrimination is \cite{helstrom}
  \be
 \epsilon_{\rm eq}= \frac{1}{2}-\frac{1}{2}
  \sqrt{1-|\langle e_{0,0} | e_{1,1} \rangle |^{2N}} ,
  \label{e}
  \ee
Her guess for Alice's symbol is denoted by $s_E$. On the other
hand, if Eve obtains the outcome corresponding to $F_{\rm dif}$,
the state of the three parties is
 \be
 [01]_{AB}\otimes [e_{0,1}]_E^{\otimes N}
 +[10]_{AB}\otimes [e_{1,0}]_E^{\otimes N} .
  \label{AdifB}
 \ee
The corresponding error probability $\epsilon_{\rm dif}$ is the
same as in Eq. (\ref{e}), with the replacement $ |\langle e_{0,0}
| e_{1,1} \rangle | \to |\langle e_{0,1} | e_{1,0} \rangle |$.
Note that $|\langle e_{0,0} | e_{1,1} \rangle | \geq |\langle
e_{0,1} | e_{1,0} \rangle |$. Eve's information now consists of
$s_E$, as well as the outcome of the measurement~\w{projectors2},
$r_E=\{{\rm eq},{\rm dif}\}$. It is shown in what follows that the
corresponding probability distribution $P(s_A,s_B,(s_E,r_E))$
cannot be distilled using one-way communication. In order to do
that, we show that Eve can always map $P$ into a new probability
distribution, $Q$, which is not one-way distillable. Therefore,
the non-distillability of $P$ is implied.

Eve's mapping from $P$ to $Q$ works as follows: she increases her
error until $\epsilon_{\rm dif} = \epsilon_{\rm eq}$. She achieves
this by changing with some probability the value of $s_E$ when
$r_E=\rm dif$. After this, Eve forgets $r_E$. The resulting
tripartite probability distribution $Q$ satisfies $Q({s_B,
s_E|s_A})=Q({s_B|s_A})\, Q({s_E|s_A})$. Additionally, we know that
$Q({s_B|s_A})$ and $Q({s_E|s_A})$ are binary symmetric channels
with error probability $\epsilon_B$($=\epsilon_N$ in \w{epN}) and
$\epsilon_{\rm eq}$ in~\w{e}, respectively. It is proven in
\cite{Maurer CAD} that in such situation the one-way key rate is
  \be
  K_{\rightarrow}= h(\epsilon_{\rm eq})-h(\epsilon_B),
  \label{ratel2}
  \ee
which is non-positive if
  \be
  \epsilon_{\rm eq} \leq \epsilon_B \ .
  \label{be}
  \ee
Let us finally show that this inequality is satisfied for all
values of $N$ whenever the condition \w{security} does not hold.
Writing $z=\lambda_1+\lambda_2$, we have $1/2\leq z\leq1$, since
the state of Alice and Bob is assumed entangled. Using the
following inequality
  \bea
  \frac{1}{2}-\frac{1}{2}\sqrt{1-\left( \frac{1-z}{z} \right)^N} \leq
  \frac{(1-z)^N}{z^N+(1-z)^N} ,\label{prev}
  \eea
which holds for any positive $N$, the right-hand side of~\w{prev}
is equal to $\epsilon_B$, whereas the left-hand side is an upper
bound for $\epsilon_{\rm eq}$. This bound follows from the
inequality $(\lambda_1-\lambda_2)^2/z^2 \leq (1-z)/z$, which is
the negation of \w{security}. That is, if condition \w{security}
is violated, no secret key can be distilled with our SIMCAP
protocol. More precisely, there exists no $N$ such that CAD
followed by one-way distillation allows to establish a secret key.
Since \w{security} is sufficient for security, the attack we have
considered is in some sense optimal and the security bound
\w{security} is tight for our SIMCAP protocol.

It is worth analyzing the resources that this optimal
eavesdropping attack requires. First of all, note that Eve does
not need to perform any coherent quantum operation, but she only
requires single-copy level (individual) measurements. This is
because when discriminating $N$ copies of two states, there exists
an adaptative sequence of individual measurements which achieves
the optimal error probability \w{e} \cite{2-state discrimination}. However, what
Eve really needs is the ability to store her quantum states after
listening to the (public) communication exchanged by Alice and Bob
during the CAD part of the protocol.

\subsection{Inequivalence of CAD1 and CAD2 for individual attacks}
\label{secineq}

As we have seen, the two CAD protocols lead to the same security
condition. This follows from the fact that Eve is not assumed to
measure her state before the CAD takes place. Then, she can
effectively map one CAD protocol into the other by means of the
reversible operation $U_E$. This is no longer true in the case of
individual attacks. Interestingly, in this scenario, the two
two-way distillation methods do not give the same security
condition. As mentioned, although the study of individual attacks
gives a weaker security, it is relevant in the case of realistic
eavesdroppers. Moreover, we believe the present example has some
interest as a kind of toy model illustrating the importance of the
reconciliation part for security. Recall that in the case of
individual attacks, where Eve can neither perform coherent
operations nor have a quantum memory, the security condition using
$CAD2$ is the entanglement condition $\lambda_1>1/2$
\cite{equivalence in 2}. However, when the honest parties apply
$CAD1$ plus one-way communication, the security condition is
\w{security}. This holds true for two-qubit protocols, and
remains open for the two-qudit protocols studied in the next sections \cite{open problem}. 

Let us suppose that Alice and Bob apply $CAD1$ and consider the
following individual attack. Eve knows that for all the instances
passing the CAD protocol, Alice and Bob's symbols are equal with
very high probability. Moreover, she knows that in all the
position announced by Alice, Alice's symbol is the same.
Therefore, from her point of view, the problem reduces to the
discrimination of $N$ copies of the two states $\ket{e_{i,i}}$.
Thus, she has to apply the measurement that optimally
discriminates between these two states. As mentioned, the optimal
two-state discrimination \cite{2-state discrimination} can be
achieved by an adaptive individual measurement strategy.
Therefore, Eve can apply this adaptive strategy to her states
right after her individual interaction. Her error probability is
again given by (\ref{e}). That is, although the attack is
individual, the corresponding security condition is the same as
for collective attacks.

This $N$-copy situation on Eve's space does not happen when Alice
and Bob apply $CAD2$. Indeed, Eve maps $CAD2$ into $CAD1$ by
applying the correcting unitary operation $U_i$ after knowing the
vector $X$ used in $CAD2$. This is the key point that allowed her
to map one situation into the other above. This is however not
possible in the case of individual attacks, where Eve is assumed
to measure before the reconciliation part takes place. Under
individual attacks, the security condition for $CAD2$ is
equivalent to the entanglement condition for Bell diagonal states,
as shown in \cite{equivalence in 2}. Therefore, the two CAD
protocols, which have proven to be equivalent in terms of
robustness against general quantum attacks, become inequivalent in
the restricted case of individual attacks.

\section{BB84 and six-state protocols}

The goal of the previous study has been to provide a general
formalism for determining the security of qubit channels under a
class of realistic QKD protocols. Relevant \emph{prepare and
measure} schemes, such as the BB84 and six-state protocol,
constitute a particular case of our analysis. Indeed, the process
of correlation distribution and channel tomography in these
protocols is done by Alice preparing states from and Bob measuring
in two (BB84) or three (six-state) bases. In this section, we
apply the derived security condition to these protocols and
compare the obtained results with previous security bounds. As
explained in \ref{revbounds}, a standard figure of merit in the
security analysis of a given QKD protocol is given by the maximum
error rate such that key distillation is still possible. For
instance, in the case of one-way communication, the values of the
critical error rates keep improving (see \cite{SRS} for the latest
result in this sense) since the first general security proof by
Mayers \cite{Mayers}. In the case of reconciliation using two-way
communication, the best known results were obtained by Chau in
\cite{chau}. It is then important to know whether these bound can
be further improved. In what follows, it is shown that our
necessary condition for security implies that Chau's bounds cannot
be improved. In order to do that, then, one has to employ other
reconciliation techniques, different from advantage distillation
plus one-way standard techniques. Some of these possibilities are
discussed in the next sections.

\subsection{BB84 protocol}

In the BB84 protocol \cite{BB84}, bits are encoded into two sets
of mutually unbiased bases $\{|0\rangle, | + \rangle\}$ and
$\{|1\rangle, |-\rangle \}$ respectively, where $|\pm\rangle =
(|0\rangle \pm |1\rangle)/\sqrt{2}$. One can easily see that in
the entanglement-based scheme, a family of attacks by Eve
producing a QBER $Q$ is given by the Bell-diagonal states (see
also \cite{CRE}) \be \rho_{AB} = (1-2Q+x)[\Phi_{1}]+
(Q-x)[\Phi_{2}] + (Q-x)[\Phi_{3}] + x [\Phi_{4}],
\label{bb84state} \ee since the QBER is \bea Q & = & \langle 01
|\rho_{AB}|01\rangle + \langle 10
|\rho_{AB}|10\rangle \nonumber \\
& = & \langle +- |\rho_{AB}|+-\rangle + \langle -+
|\rho_{AB}|-+\rangle \label{qber BB84}\eea and $0\leq x\leq Q$.
When Alice and Bob apply one-way communication distillation, the
attack that minimizes (\ref{dw}) is $x=Q^{2}$, and leads to the
well-known value of $QBER=11\%$, first obtained by Shor and
Preskill in \cite{Shor and Preskill}. The corresponding unitary
interaction by Eve is equal to the phase-covariant cloning
machine, that optimally clones qubits in an equator (in this case,
in the $xz$ plane).

When one considers the two-way distillation techniques studied in
this work, condition (\ref{security}), or (\ref{security2}),
applies. Then, one can see that the optimal attack, for fixed
QBER, consists of taking $x=0$. Therefore, Eve's attack is, not
surprisingly, strongly dependent on the type of reconciliation
employed. In the case of two-way communication, Eve's optimal
interaction can also be seen as a generalized phase-covariant
cloning transformation, which is shown in the Appendix I. Using
this attack, the derived necessary condition for security is
violated when $QBER=20\%$. This is precisely the same value
obtained by Chau in his general security proof of BB84
\cite{chau}. So, the considered collective attack turns out to be
tight, in terms of robustness. Recall that the security bound
against individual attacks is at the entanglement limit, in this
case giving $QBER=25.0 \%$ \cite{equivalence in 2,ppt}. The full
comparison is depicted in the Fig. \w{two-way}.

Note also that the state (\ref{bb84state}) with $x=0$, associated
to the optimal attack, does not fit into our canonical form for
Bell diagonal states, since $\lambda_{2}$ is not the minimal Bell
coefficient. This simply means that key distillation from this
state using a SIMCAP protocol is still possible. Alice and Bob
only have to measure in a different basis, namely in the $y$
basis. That is, if Alice and Bob knew to share this state, or
channel, and could prepare and measure states in the $y$ basis,
not used in the considered version of BB84, they would be able to
establish a secure key. This channel is still useful for QKD using
a prepare and measure scheme, although not using the considered
version of BB84. In our opinion, this illustrates why the present
approach, that aims at identifying secrecy properties of channels
without referring to a given protocol, is more general.

\subsection{Six-state protocol}

If a third mutually unbiased basis, in the $y$ direction, is added
to BB84, one obtains the so-called six-state protocol. The
information encoding is as follows: bit 0 is encoded on states
$\{|0\rangle, |+\rangle, |+ i\rangle\}$, and $1$ in $\{|1\rangle,
|-\rangle, |- i\rangle\}$, where $|\pm i \rangle = (|0\rangle \pm
i |1\rangle)/\sqrt{2}$ \cite{6 state}. It is easy to see that an
attack by Eve producing a QBER equal to $Q$ is given by the Bell
diagonal state
\begin{equation}\label{wstates}
    \rho_{AB} =
(1-\frac{3}{2}Q)[\Phi_{1}]+ \frac{Q}{2}[\Phi_{2}]
+\frac{Q}{2}[\Phi_{3}]+\frac{Q}{2}[\Phi_{4}].
\end{equation}
This attack actually corresponds to Eve applying the universal
cloning transformation. Contrary to what happened for BB84, this
attack is optimal for both types of reconciliation protocols,
using one- or two-way communication.

Applying the security condition \w{security}, the security bound
gives a critical QBER of $Q=27.6 \%$. This value again coincides
with the one obtained by Chau in his general security proof of
\cite{chau} for the six-state protocol. The present attack, then,
is again tight. In the case of indidual attacks, the security
bound \cite{equivalence in 2} is the entanglement limit
$Q=33.3\%$.

\begin{figure}
  \includegraphics[width=8cm]{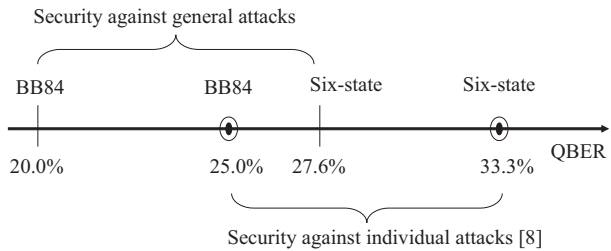}\\
  \caption{Security bounds of the BB84 and the six-state protocols
  against individual and collective attacks: When Eve is supposed
  to apply individual attacks, all entangled states are
  distillable to a secret key. Assuming general attacks, security
  bounds are $20.0 \%$ and $27.6 \%$, respectively, for the BB84 and
  the six-state protocols. This means that non-distillable secret
  correlations may exist(see, the section VI).
  }\label{bounds}
\end{figure}

\section{Can these bounds be improved?}

The previous section has applied the obtained security condition
to two well-known QKD protocols. In the corresponding attack, Eve
is forced to interact individually and in the same way with the
sent qubits. As discussed, the de Finetti results by Renner imply
that this does not pose any restriction on Eve's attack. However,
Eve is also assumed to measure her states right after CAD, while
she could have delayed her measurement, for instance until the end of the
entire reconciliation. In spite of this apparent limitation, the
condition is shown to be tight, under the considered distillation
techniques, for the two protocols. As it has been mentioned, the
obtained bounds do not coincide with the entanglement limit. This
raises the question whether prepare and measure schemes, in
general, do attain this limit. Or in other words, it suggests the
existence of channels that, although can be used to distribute
distillable entanglement, are useless for QKD using prepare and
measure techniques. Recall that a channel that allows to establish
distillable entanglement is secure: this just follows from
combining the de Finetti argument with standard entanglement
distillation. So, in this sense the channel indeed contains
distillable secrecy. However, our results suggest that this
secrecy is non-distillable, or bound, using single-copy
measurements. That is, this secrecy is distillable only if both
parties are able to perform coherent quantum operations. Perhaps,
the simplest example of this channel is given by (\ref{wstates})
with $Q>27.6\%$, i.e. by a weakly entangling depolarizing channel.

The aim of this section is to explore two possibilities to improve
the previous security bounds. We first consider the classical
pre-processing introduced in \cite{RNB}. In this work, previous
security bounds using one-way communication protocols for BB84 and
six-state protocols have been improved by allowing one of the
honest parties to introduce some local noise. This noise worsens
the correlations between Alice and Bob, but it deteriorates in a
stronger way the correlations between Alice and Eve. Here, we
study whether a similar effect can be obtained in the case of the
considered two-way communication protocols. In a similar way as in
Ref. \cite{RNB}, we allow one of the two parties to introduce some
noise, given by a binary symmetric channel (BSC). In our case,
however, this form of pre-processing does not give any improvement
on the security bounds. Later, we study whether
the use of coherent quantum operations by one of the parties
helps. We analyze a protocol that can be understood as a hybrid
between classical and entanglement distillation protocol.
Remarkably, this protocol does not provide any improvement either.
In our opinion, these results strengthen the conjectured bound
secrecy of these weakly entangled states when using SIMCAP
protocols \cite{note prepr}.

\subsection{Pre-processing by one party}

Recently, it has been observed that local classical pre-processing
by the honest parties of their measurement outcomes can improve
the security bounds of some QKD protocols \cite{RNB}. For
instance, Alice can map her measurement values $X$ into another
random variable $U$, and this transforms the mutual information
from $I(X:B)$ into $I(U:B)$. At the same time, $I(X:E)$ changes to
$I(U:E)$. In general, this mapping makes the mutual information of
Alice and Bob decrease, but bounds on the secret key rate may
improve, e.g. $I(U:B) - I(U:E)>I(X:B) - I(X:E)$. Actually, by
applying a simple BSC of probability $q$, where the input value is
kept unchanged with probability $1-q$ or flipped with probability
$q$, Alice may be able to improve the one-way secret-key rate \cite{RNB}.
Using this technique, the security bounds have been moved from
$11\%$ to $12.4\%$ for the BB84 protocol and from $12.7\%$ to
$14.1\%$ in the six-state protocol \cite{RNB}. Here, we analyze
whether a similar effect happens in the case of protocols
consisting of two-way communication. Note that pre-processing is
useless if applied after CAD. Indeed, recall that the situation after CAD for
the attack of Section \ref{sectatt} is simply given by two
independent BSC channels between Alice and Bob and Alice and Eve,
where pre-processing is known to be useless. The only possibility
left is that Alice and/or Bob apply this pre-processing before the
whole reconciliation protocol takes place.

As mentioned, Alice's pre-processing consists of a BSC channel,
where her measurement value $j$ is mapped into $j$ and $j+1$ with
probabilities $1-q$ and $q$, respectively. After this classical
pre-processing, the state of the three parties is \bea
\sigma_{ABE} \propto \sum_{i,j}[i,j]_{AB}\otimes
[\widetilde{\rho_{i,j}}] \nonumber \eea where \bea
\label{evestates} \widetilde{\rho_{0,0}} & = & (1-q)
(1-\epsilon_{AB}) [e_{0,0}]+
q \epsilon_{AB}[e_{1,0}] \nonumber \\
\widetilde{\rho_{0,1}} & = & (1-q) \epsilon_{AB} [e_{0,1}] + q
(1-\epsilon_{AB})[e_{1,1}] \nonumber \\
\widetilde{\rho_{1,0}} & = & q (1-\epsilon_{AB}) [e_{0,0}] +
(1-q) \epsilon_{AB} [e_{1,0}] \nonumber \\
\widetilde{\rho_{1,1}} & = & q \epsilon_{AB} [e_{0,1}] +
(1-q)(1-\epsilon_{AB}) [e_{1,1}] \nonumber \\
\eea and $\epsilon_{AB}$ denotes the QBER of the original
measurement data, i.e. the error rate before applying
pre-processing. Again, the states with tilde are not normalized,
so \bea \widetilde{\rho_{i,i}} & = & \left(
(1-q)(\frac{1-\epsilon_{AB}}{2})+q\frac{\epsilon_{AB}}{2}
\right)\rho_{i,i} \nonumber \\
\widetilde{\rho_{i,i+1}} & = & \left(
(1-q)\frac{\epsilon_{AB}}{2}+q(\frac{1-\epsilon_{AB}}{2})
\right)\rho_{i,i+1} . \nonumber \eea Next, Alice and Bob apply
two-way CAD to $\sigma_{ABE}^{\otimes N}$. A new
error rate is obtained after CAD. The rest of the distillation
part, then, follows the same steps as in section V-A.

\begin{figure}
  \includegraphics[width=8cm]{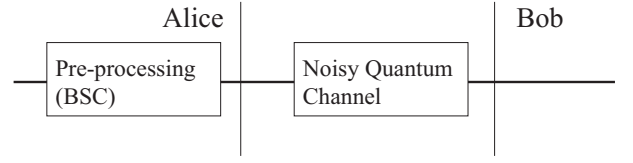}\\
  \caption{Considered classical pre-processing: Alice introduces some extra noise by permuting her classical variable with probability $q$.
  }\label{pre-processing}
\end{figure}

We now compute the mutual information between the honest parties
after CAD. The new error rate of Alice and Bob is introduced by
the BSC above, and is expressed as $\omega =
tr_{ABE}[\sigma_{ABE}(|01 \rangle_{AB}\langle 01|+|10
\rangle_{AB}\langle 10|)] = (1-q)\epsilon_{AB}
+q(1-\epsilon_{AB})$. For large $N$, the mutual information of
Alice and Bob tends to, c.f. (\ref{largeN}),

\bea I^{P}(A:B) \approx 1+ (\frac{\omega}{1-\omega})^{N} \log
(\frac{\omega}{1-\omega})^{N}. \nonumber \eea In the same limit,
Eve's state can be very well approximated by \bea \sigma_{E}
\approx \frac{1}{2}(\rho_{00}^{\otimes N} + \rho_{11}^{\otimes N})
\nonumber, \eea since $||\widetilde{ \rho_{i,i} }||
> ||\widetilde{ \rho_{i,j} ||}$. After some patient algebra,
one can see that the Holevo information of Alice and Eve channel
is (see also Appendix II): \bea I^{P}(A:E) \approx 1 -
\frac{1}{\ln 4}(u|\langle e_{0,0}\ket{e_{1,1}}|^2+v|\langle
e_{0,1}\ket{e_{1,0}}|^2)^N \nonumber \eea where \bea u =
\frac{(1-q)(1-\epsilon_{AB})}
{q\epsilon_{AB}+(1-q)(1-\epsilon_{AB})}, \nonumber \eea and
$u+v=1$. The case of $q=0$ (or equivalently, $u=1$) recovers the
initial mutual information $I(A:E)$. Therefore, the security
condition of this protocol is

\bea \label{keycondpr}
u|\bra{e_{0,0}}e_{1,1}\rangle|^{2}+v|\bra{e_{0,1}}e_{1,0}\rangle|^{2}
>  \frac{\omega}{1-\omega}.\eea
More precisely, whenever this condition is satisfied, there exists
a finite $N$ such that $I^{P}(A:B) - I^{P}(A:E)> 0 $.

The derived bound looks again intuitive. The r.h.s quantifies how
Alice and Bob's error probability for the accepted symbols
converges to zero when $N$ is large. If one computes the trace
distance between $\rho_{0,0}$ and $\rho_{1,1}$, as defined in Eq.
(\ref{evestates}), one can see that
\begin{equation}
    \tr|\rho_{0,0}-\rho_{1,1}|\approx
    2-(u|\bra{e_{0,0}}e_{1,1}\rangle|^2+v|\langle
e_{0,1}\ket{e_{1,0}}|^2)^N ,
\end{equation}
which gives the l.h.s. of (\ref{keycondpr}). This result suggests
that the derived condition may again be tight. That is, it is
likely there exists an attack by Eve breaking the security of the
protocol whenever (\ref{keycondpr}) is not satisfied. This attack
would basically be the same as above, where Eve simply has to
measure after the CAD part of the protocol.

Our goal is to see whether there exist situations where
pre-processing is useful. Assume this is the case, that is, there
exists a state for which (\ref{keycondpr}) holds, for some value
of $q$, while (\ref{security}) does not. Then,
\begin{equation}
\frac{\epsilon_{AB}}{1-\epsilon_{AB}} \geq |\langle e_{00}|e_{11}
\rangle|^{2} > \frac{1}{u}(\frac{\omega}{1-\omega} - v|\langle
e_{01}|e_{10} \rangle|^{2}).
\end{equation} After some simple algebra, one gets the
inequality: \bea \frac{1}{\epsilon_{AB}} < 1 + |\langle
e_{01}|e_{10} \rangle |^{2}. \nonumber \eea The r.h.s. of this
equation is smaller than 2, and this implies that $\epsilon_{AB}
> 1/2$. However, this contradicts $0\leq \epsilon_{AB} < 1/2$,
so we conclude that one-party pre-processing does not improve the
obtained security bound.

Notice that since the reconciliation part uses communication in
both directions, it seems natural to consider pre-processing by
the two honest parties, where Alice and Bob introduce some noise,
described by the probabilities $q_A$ and $q_B$. In this case,
however, the analytical derivation is much more involved, even in
the case of symmetric pre-processing. Our preliminary numerical calculations
suggest that two-parties pre-processing may be useless as well.
However, these calculations should be interpreted in a very
careful way. Indeed, they become too demanding already for a
moderate $N$, since one has to compute the von Neumann entropies
for states in a large Hilbert space, namely $\rho_{0,0}^{\otimes
N}$ and $\rho_{1,1}^{\otimes N}$. Therefore, the detailed analysis of pre-processing by the two honest parties remains to be done.

Before concluding, we would like to mention that pre-processing,
before or after CAD, may help in improving the distillable
secret-key rate if the initial rate without pre-processing is
already positive (see for instance \cite{Renner}). However, this
improvement vanishes for large blocks and the obtained security
bounds do not change.

\subsection{Bob's coherent operations
do not improve the security bound}

In order to improve the security bound, we also consider the
scenario where Bob performs some coherent quantum operations
before his measurement. Thus, he is assumed to be able to store
quantum states and manipulate them in a coherent way, see Fig. 8.
This is very unrealistic, but it gives the ultimate limit for
positive key-rate using the corresponding prepare and measure
protocol. We do not solve the problem in full generality. Here we
consider the rather natural protocol where Bob applies the
recurrence protocol used in entanglement distillation. That is, he
applies CNOT operations to $N$ of his qubits and measures all but
one. He accepts only when the results of these $N-1$ measurements
are zero and keeps the remaining qubit. Later Bob applies a collective measurement on all the accepted qubits. Alice's part of the protocol remains unchanged.

After Alice has measured her states and announced the position of
$N$ symbols having the same value, Alice-Bob-Eve state reads
\begin{equation}\label{cqqcorr}
    \rho_{ABE}=[0]_A\otimes[\widetilde{be_0}]_{BE}^{\otimes N}+
    [1]_A\otimes[\widetilde{be_1}]_{BE}^{\otimes N} ,
\end{equation}
where $\ket{\widetilde{be_i}}=\langle i\ket{\psi}_{ABE}$. Note that Alice, Bob
and Eve now share CQQ correlations. Bob applies his part of the
protocol and accepts. The resulting state turns out to be equal
to, up to normalization,
\begin{eqnarray}
    \rho_{ABE}^N&\propto[0]\otimes[\ket 0\ket{\widetilde{e_{0,0}}}^{\otimes N}+
    \ket 1\ket{\widetilde{e_{0,1}}}^{\otimes N}]+\nonumber\\
    &[1]\otimes[\ket 0\ket{\widetilde{e_{1,0}}}^{\otimes N}+
    \ket 1\ket{\widetilde{e_{1,1}}}^{\otimes N}] .
\end{eqnarray}
Since Bob is allowed to apply any coherent operation, the
extractable key rate satisfies (\ref{dw}), where now both
information quantities, $I(A:B)$ and $I(A:E)$, are equal to the
corresponding Holevo bound. Of course $I(A:E)$ has not changed. It is straightforward to see that one
obtains the same bound for the key rate as for the state (\ref{ccqcorr}). This follows from the fact that $\langle e_{i,i}\ket{e_{i,j}}=0$, where $i\neq j$. Then, this hybrid protocol does not provide any advantage with respect to
SIMCAP protocols.

\begin{figure}
\label{qadfig}
  \includegraphics[width=8cm]{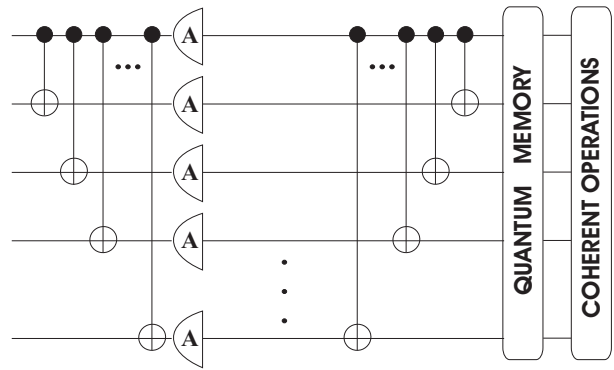}\\
  \caption{Quantum advantage distillation protocol: Alice performs single-copy measurement and processes the obtained classical outcomes.
Bob keeps his quantum states on a quantum memory and performs coherent quantum operations.}\label{QAD}
\end{figure}

Recall that if the two parties apply coherent quantum operations,
they can run entanglement distillation and distill from any
entangled two-qubit state. Actually a slightly different protocol
where (i) both parties perform the coherent recurrence protocol
previously applied only by Bob, (ii) measure in the computational
bases and (iii) apply standard one-way reconciliation techniques
is secure for any entangled state. As shown, if one of the parties
applies the ``incoherent" version of this distillation protocol,
consisting of first measurement and later CAD, followed by
classical one-way distillation, the critical QBER decreases.

\section{Generalization to arbitrary dimension}
\label{secqudits}

In the previous sections we have provided a general formalism for
the study of key distribution through quantum channels using
prepare and measure schemes and two-way key distillation. In the
important case of Pauli channels, we have derived a simple
necessary and sufficient condition for security, for the
considered protocols. In the next sections, we move to higher
dimension, where the two honest parties employ $d-$dimensional
quantum systems, or qudits. The generalization of the previous
qubit scenario to arbitrary dimension is straightforward. Alice
locally generates a $d-$dimensional maximally entangled state,
\bea \ket{\Phi}= \frac{1}{\sqrt{d}} \sum_{k=0}^{d-1}|k\rangle
|k\rangle \label{d-bell} \eea measures the first particle of the
pair, and sends the other one to Bob. Since the channel between
Alice and Bob is noisy, the shared state will change into a mixed
state $\rho_{AB}$. As usual, all the noise in the channel is due
to Eve's interaction.

In what follows, we consider generalized Pauli channels. For these
channels, Eve introduces flip and phase errors, generalizing the
standard bit-flip $\sigma_{x}$ and phase-flip $\sigma_{z}$
operators of qubits. This generalization is given by the unitary
operators \bea U_{m,n}= \sum_{k=0}^{d-1} \exp(\frac{2 \pi i
}{d}kn)|k+m\rangle \langle k |.\nonumber \eea Thus, a quantum
system in state $\rho$ propagating through a generalized Pauli
channel is affected by a $U_{m,n}$ flip with probability
$p_{m,n}$, that is \be  D(\rho) = \sum_{m,n} p_{m,n} U_{m,n}\rho
U_{m,n}^{\dagger}\nonumber .\ee When applied to half of a
maximally entangled state $\ket{\Phi}$, the resulting state is
Bell-diagonal, \bea (\one\otimes D)(\Phi)=
\sum_{m=0}^{d-1}\sum_{n=0}^{d-1} p_{m,n} |B_{m,n}\rangle \langle
B_{m,n} |, \label{d-bell diagonal} \eea where the states
$\ket{B_{m,n}}$ define the generalized Bell basis
\begin{equation}\label{Bellbasis}
    \ket{B_{m,n}}=(\one\otimes U_{m,n})\ket{\Phi}=\frac{1}{\sqrt d}
    \sum_{k=0}^{d-1}e^{\frac{2 \pi i
}{d}kn}\ket k\ket{k+m} .
\end{equation}
The global state including Eve reads
\begin{equation}
\label{psiabe}
    \ket{\psi_{ABE}}=\sum_{m=0}^{d-1}\sum_{n=0}^{d-1} c_{m,n}
    \ket{B_{m,n}}_{AB}\ket{m,n}_E ,
\end{equation}
where $c_{m,n}^2=p_{m,n}$ and $\{\ket{m,n}\}$ defines a basis.

In the next lines, we derive a security conditions for these
channels when the two honest parties measure in the computational
bases. We restrict to the computational bases for the sake of
simplicity, although the main ideas of the formalism can be
applied to any bases, and then numerically optimized. We then
generalize the previous eavesdropping attack. Contrary to what
happened in the qubit case, we are unable to prove the tightness
of our condition in full generality using this attack.

We then apply the derived security condition to the known
protocols in $d$-dimensional systems, such as the $2$- and
$(d+1)$-bases protocols. These protocols can be seen as the
natural generalization of the BB84 and the six-state protocols to
higher dimension \cite{cbkg}. Exploiting the symmetries of these
schemes, we can prove the tightness of our security condition for
these protocols. In the case of the $(d+1)$-bases protocol, some
security bounds using two-way communication have been obtained by
Chau in \cite{chaud}. Here, we obtain the same values, therefore
proving that they cannot be improved unless another reconciliation
protocol is employed. Moreover, in the case of $2$-bases protocol, we
derive the same security bound as in \cite{GKG}. Thus, again,
another reconciliation protocol is necessary if the bound is to be
improved.


\subsection{Sufficient condition}

After sending half of a maximally entangled state through the
Pauli channel, Alice and Bob share the state \bea \rho_{AB} =
\sum_{m,n} p_{m,n} |B_{m,n}\rangle \langle B_{m,n}| , \nonumber
\eea where the probabilities $p_{m,n}$ characterize the
generalized Pauli channel. After measuring in the computational
bases, the two honest parties obtain correlated results. We denote
by $F$, fidelity, the probability that Alice and Bob get the same
measurement outcome. It reads \bea F & = & \sum_{k=0}^{d-1}\langle
kk |\rho_{AB}|kk\rangle = \sum_{n}p_{0,n}.\nonumber \eea In a
similar way as for the qubit case, we introduce a measure of
disturbance for the $d-1$ possible errors. Denote Alice's
measurement result by $\alpha$. Then, Bob obtains $\alpha+j$, with
probability \be D_{j}  = \sum_{\alpha=0}^{d-1}P(A=\alpha,B=\alpha
+j)=\sum_{n=0}^{d-1} p_{j,n}. \nonumber \ee The total disturbance
is defined as \be D = \sum_{j\neq 0} D_{j} . \ee Of course,
$D_0=F$. Notice that all the $D_{j}$ can be taken smaller than
$F$, without loss of generality. Indeed, if this was not the case,
the two honest parties could apply local operations $U_{m,n}$ to
make the fidelity $F$ larger than any other $D_{j}$. Note also
that the errors have different probabilities  $D_{j}$.

We now include Eve in the picture, the resulting global state
being (\ref{psiabe}). As for the qubit case, Eve's interaction by
means of the Pauli operators can be formulated as an asymmetric
$1\rightarrow 1+1$ cloning transformation \cite{asyclon}. In what
follows, and again invoking the de Finetti argument, it is assumed
that Alice, Bob and Eve share many copies of the state
(\ref{psiabe}). After the measurements by Alice and Bob, the
quantum state describing the CCQ correlations between the three
parties is \bea \rho_{ABE} \propto \sum_{\alpha=0}^{d-1}
\sum_{\beta=0}^{d-1} [\alpha,\beta]_{AB} \otimes
[\widetilde{e_{\alpha, \beta}}]_{E}. \label{abed}\eea
Eve's states are \bea |e_{\alpha,\alpha}\rangle & = &
\frac{1}{\sqrt{F}} \sum_{n=0}^{d-1} c_{0,n}e^{\frac{2\pi
i}{d}\alpha
n}|0,n\rangle \nonumber \\
|e_{\alpha,\beta}\rangle & = & \frac{1}{\sqrt{D_{\beta-\alpha}}}
\sum_{n=0}^{d-1} c_{\beta-\alpha,n}e^{\frac{2\pi i}{d}\alpha
n}|\beta-\alpha,n\rangle \nonumber \\
\eea where the algebra is modulo $d$ and $\beta\neq \alpha$. As
above, the states with tilde are not normalized, \bea
|\widetilde{e_{\alpha, \alpha}}\rangle & = & \sqrt{F}
|e_{\alpha, \alpha}\rangle \nonumber \\
|\widetilde{e_{\alpha, \beta}}\rangle & = &
\sqrt{D_{\beta-\alpha}} |e_{\alpha, \beta}\rangle . \nonumber
\eea Note that $\langle e_{\alpha,\beta}|e_{x,y}\rangle = 0$
whenever $\beta-\alpha \neq y-x$, so Eve can know in a
deterministic way which error (if any) occurred between Alice and
Bob.

After the measurements, Alice and Bob have a list of correlated
measurement outcomes. They now apply CAD. First, Alice locally
generates a random variable, $s_A$, that can take any value
between 0 and $d-1$ with uniform probability. She then takes $N$
of her symbols $(\alpha_{1},\cdots, \alpha_{N})$ and announces the
vector $\vec X=(X_{1},\cdots,X_{N})$ such that
$X_{j}=s-\alpha_{j}$. Bob sums this vectors to his corresponding
symbols $(\beta_{1},\cdots, \beta_{N})$. If the $N$ results are
equal, and we denote by $s_B$ the corresponding result, he accepts
$s_B$. It is simple to see that Bob accepts a symbol with
probability $p_{ok}=F^{N} + \sum_{j=1}^{d-1}D_{j}^{N}$. After
listening to the public communication used in CAD, Eve knows
$(X_{1},\cdots ,X_{N})$. As in the previous qubit case, she
applies the unitary operation: \bea \mathcal{U}_{E} & = & \sum_{m
= 0}^{d-1} \sum_{l = 0}^{d-1} e^{\frac{2\pi i }{d}X_{j} m} [
l,-m]\label{ud}\eea This unitary operation transforms
Eve's states as follows, \bea \mathcal{U}_{E}^{\otimes N}
:\bigotimes_{j=0}^{N}|e_{\alpha_{j},\beta_{j}}\rangle
\longrightarrow
 \bigotimes_{j=0}^{N}|e_{s,s-(\alpha_{j}-\beta_{j})} \rangle . \nonumber  \eea
As above, this operation makes Alice, Bob and Eve's state
independent of the specific vector used for CAD. The resulting
state reads
 \be \sum_{s_A,s_B=0}^{d-1}
 [s_A,s_B]_{AB}\otimes [e_{s_A,s_B}]_E^{\otimes N},
  \label{AeqB1}
 \ee
up to normalization. As above, the goal is to see when it is
possible to find a finite $N$ such that the CCQ correlations of
state (\ref{AeqB1}) provide a positive key-rate, according to the
bound of Eq. (\ref{dw}).

The new disturbances $D'_j$, $j=1,\ldots,d-1$, after the CAD
protocol are equal to
\begin{equation}\label{distad}
    D'_j=\frac{D_j^N}{\sum_{k=0}^{d-1}D_k^N}\leq
    \left(\frac{D_j}{F}\right)^N ,
\end{equation}
where, again, the last inequality tends to an equality sign for
large $N$. The mutual information between Alice and Bob is \be
I(A:B) = \log d +
\frac{F^N}{p_{ok}}\log\frac{F^N}{p_{ok}}+\sum_{j=1}^{d-1}D'_{j}\log
D'_{j} .\ee For large $N$, this quantity tends to \bea I(A:B) & =
& \log d - N\left(\frac{D_{m}}{F}\right)^{N}\log \frac{F}{D_{m}} +
O((\frac{D_{m}}{F})^{N}) \nonumber \eea where $D_{m}=\max_j D_j$
for $j\in\{1,\cdots,d-1\}$.

Let us now compute Eve's information. Again, since Alice and Eve
share a CQ channel, Eve's information is measured by the Holevo
bound. For very large $N$, as in the case of qubits, we can
restrict the computation of $\chi(A:E)$ to the cases where there
are no errors between Alice and Bob after CAD. So, Eve has to
distinguish between $N$ copies of states $\ket{e_{k,k}}$. Thus, in
this limit, $\chi(A:E)\approx S(\rho_E)$, where
\begin{equation}
    \rho_E=\frac{1}{d}\sum_k [e_{k,k}^{\otimes N}] .
\end{equation}
Denote by $A_{\eta}$, with $\eta=0,\ldots,d-1$, the eigenvalues of
$\rho_E$. As shown in Appendix III, one has \bea A_{\eta} =
\frac{1}{d^{2}} \sum_{k=0}^{d-1}\sum_{k^{'}=0}^{d-1} e^{\frac{2\pi
i}{d}\eta (k-k^{'})}\langle e_{k}|e_{k^{'}} \rangle^{N}. \nonumber
\eea
Decomposing the eigenvalue $A_{\eta}$ into the term with $k=k^{'}$
and with $k\neq k^{'}$, we can write $A_{\eta} = (1+
X_{\eta}^{(N)}/d)/d$, where
\begin{equation}\label{ }
    X_{\eta}^{(N)} =  \sum_{k\neq k^{'}} e^{\frac{2\pi i}{d}\eta
(k-k^{'})}\langle e_{k,k}|e_{k^{'},k^{'}} \rangle^{N}.
\end{equation}
Note that $X_{\eta}^{(N)}$ is real since $X_{\eta}^{(N)} =
d^{2}A_{\eta} -d$ and $A_{\eta}$ is real, and $\sum_{\eta =
0}^{d-1} X_{\eta}^{(N)} = 0$ because of normalization. Moreover,
$X_{\eta}^{(N)}$ goes to zero when $N$ increases. Using the
approximation $\log(1+x)\approx x/\ln 2 $ valid when $x\ll 1$, we
have \bea \chi(A:E) & \approx & -\sum_\eta A_\eta\log
A_\eta\nonumber\\&\approx&\log d - \frac{1}{d^{3}\ln
2}\sum_{\eta=0}^{d-1} X_{\eta}^{(N)} X_{\eta}^{(N)}
\nonumber \\
& = & \log d - \frac{d-1}{d\ln 2}\sum_{k\neq k^{'}} |\langle
e_{k,k}|e_{k^{'},k^{'}} \rangle |^{2N}. \nonumber \eea

As above, the security condition follows from the comparison of
the exponential terms in the asymptotic expressions $I(A:B)$ and
$\chi(A:E)$, having \bea \max_{k\neq k^{'}} |\langle e_{k,k} |
e_{k^{'}, k^{'}} \rangle |^{2} > \max_{j} \frac{D_{j}}{F}.
\label{security for general bell diagonal state} \eea This formula
constitutes the searched security condition for generalized Bell
diagonal states. Whenever (\ref{security for general bell diagonal
state}) is satisfied, there exists a finite $N$ such that the
secret-key rate is positive. In the next section, we analyze the
generalization of the previous attack for qubits to arbitrary
dimension.

\subsection{Eavesdropping attack}
\label{sectattackd}

We consider here the generalization of the previous qubit attack
to arbitrary dimension. Unfortunately, we are unable to use this
attack to prove the tightness of the previously derived condition,
namely Eq. (\ref{security for general bell diagonal state}), in
full generality. 
However, the techniques developed in this section can be applied
to standard protocols, such as the 2- and $d+1$-bases protocol.
There, thanks to the symmetries of the problem, we can prove the
tightness of the security condition.

The idea of the attack is the same as for the case of qubits. As
above, Eve measures after the CAD part
of the protocol. 
She first performs the $d$-outcome measurement defined by the
projectors
  \be
  M_{\rm eq} = \sum_{n} [{0,n}] ,\quad
  M_{j} = \sum_{n} [{j,n}],
  \label{projectors}
  \ee
where $j\neq 0$. The outcomes of these measurement are
denoted by $r_E$. Using this measurement Eve can know in a
deterministic way the difference between Alice and Bob's
measurement outcomes, $s_A$ and $s_B$. If Eve obtains the outcome
corresponding to $M_{\rm eq}$, she knows the tripartite state is
(up to normalization)
 \be
 \sum_{x=0}^{d-1}[xx]_{AB}\otimes [e_{xx}]_{E}^{\otimes
 N} .
  \label{AeqB}
 \ee
Now, in order to learn $s_A$, she must discriminate between the
$d$ pure states $|e_{xx}\rangle^{\otimes N}$. Due to the
symmetry of these states, the so-called square-root
measurement(SRM) \cite{HW,srm} is optimal, in the sense that it
minimizes the error probability (see Appendix IV for more
details). She then guesses the right value of $s_A$ with
probability \bea P_{{\rm eq}}^{success} & =
&\frac{1}{d^{2}}\left|~\sum_{\eta} \sqrt{\sum_{m} e^{2\pi i (\eta
m/d)}
\langle e_{m,m}|e_{0,0}\rangle^{N} }~\right|^{2} \nonumber \\
&=& \frac{1}{d^{2}} \left|~~ \sum_{\eta=0}^{d-1}\sqrt{1+
Y_{\eta}^{(N)} }~~ \right|^{2}  , \label{PeA=Bd} \eea where
\begin{equation}
    Y_{\eta}^{(N)}  =  \sum_{m=1}^{d-1} e^{\frac{2\pi i}{d}\eta m}
\langle e_{m,m}| e_{0,0}\rangle^{N},
\end{equation}
$Y_{\eta}^{(N)}$ being real. Note that $Y_{\eta}^{(N)}$ tends to
zero for large $N$. The error probability reads $\epsilon_{{\rm
eq}}=1-P_{{\rm eq}}^{success}$.

%

If Eve obtains the outcome corresponding to $M_{j}$ after the
first measurement, she knows that the three parties are in the
state (up to normalization)
\bea
\sum_{x=0}^{d-1}[x,x+j]_{AB}\otimes [e_{x,x+j}]_{E}^{\otimes N}. \nonumber \\
\eea Eve again applies the SRM strategy, obtaining
\begin{equation}
    P_{j}^{success} = \frac{1}{d^{2}} \left|~~ \sum_{\eta=0}^{d-1}\sqrt{1+
Y_{\eta}^{(j,N)} }~~ \right|^{2} ,
\end{equation}
where
\begin{equation}
    Y_{\eta}^{(j,N)}  =  \sum_{m=1}^{d-1} e^{\frac{2\pi i}{d}\eta m}
\langle e_{m,m+j}| e_{0,j}\rangle^{N},
\end{equation}
the associated error probability being
$\epsilon_j=1-P_{j}^{success}$.

As a result of this measurement, Alice, Bob and Eve share the
tripartite probability distribution $P(s_A,s_B,(s_E,r_E))$, where
$(s_E,r_E)$ represents Eve's random variables, $r_E$ ($s_E$) being
the result of the first (second) measurement. For each value of
$r_E$, Eve knows the difference between Alice and Bob's symbol and
the error in her guess for Alice's symbol. It would be nice to
relate the distillation properties of this tripartite probability
distribution to the derived security condition (\ref{security for
general bell diagonal state}), as we did in the qubit case.
Unfortunately, we are at present unable to establish this
connection in full generality. Actually, we cannot exclude that
there exists a gap for some Bell diagonal states. However, as
shown in the next section, the considered attack turns out to be
tight when applied to standard protocols, such as the 2- and
$d+1$-bases protocols.

Let us conclude with a remark on the resources Eve needs for this
attack. After applying the same unitary operation on each qudit,
Eve stores her quantum states in a quantum memory. After CAD, she
measures her corresponding block of $N$ quantum states. Recall
that in the qubit case, Eve does not need any collective
measurement, since an adaptative individual measurement strategy
achieves the fidelity of the optimal collective measurement
\cite{2-state discrimination}. In the case of arbitrary dimension,
it is unknown whether there exists an adaptative measurement
strategy achieving the optimal error probability, at least
asymptotically, when $N$ copies of $d$ symmetrically distributed
states are given \cite{open problem}.

\section{Examples : 2- and $(d+1)$-bases Protocols in Higher Dimensions}
\label{secdprot}

We now apply the previous security condition to specific protocols
with qudits, namely the so-called $2$- and $(d+1)$-bases protocols
\cite{cbkg}, which are the generalization of the BB84 and the
six-state protocols to higher dimension. In the first case, Alice
and Bob measure in two mutually unbiased bases, say computational
and Fourier transform, while in the second, the honest parties
measure in the $d+1$ mutually unbiased bases \cite{notemub}.

The optimal cloning attack for these protocols gives a Bell
diagonal state (\ref{d-bell diagonal}). However, due to the
symmetries of the protocols, the coefficients $c_{m,n}$, or
$p_{m,n}$, are such that \bea \label{cmn}
c = \left(%
\begin{array}{cccc}
  v & x & \ldots & x \\
  x & y & \ldots & y \\
  \vdots & \vdots & \ddots & \vdots \\
  x & y & \ldots & y \\
\end{array}%
\right) \eea where the normalization condition implies
$v^{2}+2(d-1)x^{2}+(d-1)^{2}y^{2}=1$. For the $d+1$-bases
protocol, which is more symmetric, one also has $x=y$.

The fidelity, that is, the probability that Alice and Bob obtain
the same outcome, is \bea F & = & \sum_{k=0}^{d-1}\langle kk
|\rho_{AB}|kk\rangle = v^{2} + (d-1)x^{2}, \nonumber \eea for all
the bases used in the protocol. The errors distribute in a
symmetric way, $D_{j}=(1-F)/(d-1)$ for all $j\neq 0$. For the
$d+1$-bases protocol, and since we have the extra constraint
$x=y$, the coefficients $c_{m,n}$ read \bea
c_{0,0} & = &  \sqrt{\frac{(d+1)F-1}{d}}  \nonumber \\
c_{m,n} & = &  \sqrt{\frac{1-F}{d(d-1)}} ~~~~{\rm for } ~ m,n\neq
0. \eea In the 2-bases protocol, $y$ is a free parameter that can
be optimized for each value of the error rate, $D$, and depending
on the reconciliation protocol. For instance, if Eve's goal is to
optimize her classical mutual information, the optimal interaction
($1\rightarrow 1+1$ cloning machine) gives (see \cite{cbkg} for
more details) \bea
c_{0,0} & = & F  \nonumber \\
c_{m,0} & = & c_{0,n} =  \sqrt{\frac{F(1-F)}{d-1}} ~~{\rm for}~ m(n)\neq 0 \nonumber \\
c_{m,n} & = & \frac{1-F}{d-1}  ~~~~{\rm for}~ m,n\neq 0 . \eea In
a similar way as in the qubit case, this choice of coefficients is
not optimal when considering two-way reconciliation protocols, as
shown in the next lines.

\subsection{Security bounds}

Having introduced the details of the protocols for arbitrary $d$,
we only have to substitute the expression of the coefficients into
the derived security condition. Because of the symmetries of the
problem, all disturbances $D_j$ and overlaps $\langle
e_{m,m}\ket{e_{0,0}}$ are equal, which means that the security
condition simply reads
\begin{equation}\label{secprot}
    |\langle e_{m,m}|e_{0,0}\rangle|^2
> \frac{D}{(d-1)F} .
\end{equation}
After patient algebra, one obtains the following security bounds:
\begin{enumerate}
    \item For $(d+1)$-bases protocol, positive key rate is possible if
        \bea
D < \frac{(d-1)(2d+1-\sqrt{5})}{2(d^{2}+d-1)}
        \eea
        The critical QBER for the 6-state protocol, $27.6\%$, is
        easily recovered by taking $d=2$.
        Recently, Chau has derived a general security proof
        for the same protocols in Ref. \cite{chaud}. Our critical values
        are the same as in his work.
    \item For the $2$-bases protocol, the critical disturbances $D$
    are
    \bea
D < \frac{(d-1)(4d-1-\sqrt{4d+1})}{2d(4d-3)}
    \eea
The optimal attack, in the sense of minimizing the critical error
rate, is always obtained for $y=0$, see (\ref{cmn}). The critical
QBER for the BB84 protocol is recovered when $d=2$. These values coincide with those obtained in \cite{GKG} for 2-bases protocols.
\end{enumerate}

Once again, there exists a gap between this security condition and
the entanglement limit. For instance, in the case of $d+1$-bases
protocols, the entanglement limit coincides with the security
condition against individual attacks \cite{equivalence in d} \bea
|\langle e_{k,k}|e_{l,l}\rangle|
> \frac{D}{(d-1)F} , \nonumber  \eea
which looks very similar to (\ref{secprot}). Thus, there exists
again weakly entangling channel where we are unable to establish a
secure key using a prepare and measure scheme.

\begin{figure}
  \includegraphics[width=8cm]{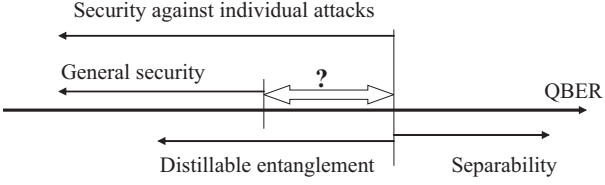}\\
  \caption{Comparison of the security bounds and the entanglement condition.
  The security condition against collective attacks requires stronger correlation than the entanglement limit.
  Again, there may exist some entangled states that are useless for key distillation with the considered techniques.}
\end{figure}

\subsection{Proof of tightness}

Finally, for these protocols, and because of the symmetries, we
are able to prove the tightness of the derived security condition,
under the considered reconciliation techniques. The goal is to
show that the probability distribution $P(s_A,s_B,(s_A,s_E))$,
resulting from the attack described in section \ref{sectattackd},
cannot be distilled using one-way communication from Alice to Bob
(the same can be proven if the communication goes from Bob to
Alice by reversing the role of these parties).

In order to do that, we proceed as in the case of qubits.
Alice-Bob's probability distribution is very simple: with
probability $F$ their symbols agree, with probability
$D_j=D/(d-1)$ they differ by $j$. After CAD on blocks of $N$
symbols, the new fidelity between Alice and Bob is
\begin{equation}
\label{fidN}
    F_N=\frac{F^N}{F^N+(d-1)\left(\frac{D}{d-1}\right)^N} .
\end{equation}
One can see that, again, Eve's error probability in guessing
Alice's symbol is larger when there are no errors between the
honest parties. As in the qubit case, Eve worsens her guesses by
adding randomness in all these cases and forgets $r_E$. After this
process, she guesses correctly Alice's symbol with probability,
see Eq. (\ref{PeA=Bd}),
\begin{eqnarray}\label{eveerrdN}
    P_{{\rm
    eq}}^{success}(N)&=&\frac{1}{d^2}\left(\sqrt{1+(d-1)\left(\frac{v-x}{F}\right)^N}
    \nonumber\right.\\
    &+&\left.(d-1)\sqrt{1+\left(\frac{v-x}{F}\right)^N}\right)^2 ,
\end{eqnarray}
independently of Bob's symbols. Here we used the fact that
$\langle e_{m,m}\ket{e_{0,0}}=(v-x)/F$ when $m\neq 0$ for the
analyzed protocols.

After Eve's transformation, the one-way distillability properties
of the final tripartite probability distribution are simply
governed by the errors, as in the qubit case. Thus, we want to
prove that at the point where the security condition is no longer
satisfied, i.e. when $((v-x)/F)^2=D/((d-1)F)$, one has
\begin{equation}
    P_{{\rm
eq}}^{success}(N)>F_N ,
\end{equation} for any block size $N$.
Define $t^2=D/((d-1)F)$, where $0\leq t\leq 1$ because $F>1/D$.
What we want to prove can also be written as, see Eqs.
(\ref{fidN}) and (\ref{eveerrdN}),
\begin{equation}
\label{condtight} \left(\frac{\sqrt{1+(d-1)t^N}
    +(d-1)\sqrt{1+t^N}}{d}\right)^2 > \frac{1}{1+(d-1)t^{2N}} ,
\end{equation}
for all $N$ and all $d$, where $0\leq t\leq 1$. Actually, using
that $0\leq t\leq 1$, it suffices to prove the case $N=1$, since
all the remaining cases will follow by replacing $t^N\to t$ and
using the condition for $N=1$. After patient algebra, one can show
that (\ref{condtight}) is satisfied for $N=1$, which finishes the
proof. Therefore, for the considered protocols, the attack
introduced above breaks the security whenever our security
condition does not hold. Therefore, this condition is tight for
the considered reconciliation techniques.

\section{Conclusion}
\label{secconcl}

This works provides a general formalism for the security analysis
of prepare and measure schemes, using standard advantage
distillation followed by one-way communication techniques. The
main tools used in this formalism are the de Finetti argument
introduced by Renner and known bounds on the key rate. We derive a
simple sufficient condition for general security in the important
case of qubit Pauli channels. By providing a specific attack, we
prove that the derived condition is tight. When applied to
standard protocols, such as BB84 and six-state, our condition
gives the critical error rates previously obtained by Chau. Since
our condition is tight, these critical error rates cannot be
improved unless another reconciliation technique is employed.
Here, most of our analysis focus on conditions for security.
However, the same techniques can be used to compute key rates.
Actually, our results imply that the critical error rates of
$20\%$ ad $27.6\%$ for the BB84 and six-state protocols can be
reached without any pre-processing by Alice, contrary to previous
derivations by Chau \cite{chau} or Renner \cite{Renner}. The rates we obtain, then, are
significantly larger. We then extend the analysis to arbitrary
dimension and generalized Bell diagonal states. The corresponding
security condition can be applied to obtain critical error rates
for the 2- and $d+1$-bases protocols. For these protocols, we can
also prove the tightness of the condition.

We explore several possibilities to improve the obtained security
bounds. As shown here, pre-processing by Alice or a coherent
version of distillation by Bob do not provide any improvement.
This is of course far from being an exhaustive analysis of all
possibilities, but it suggests that it may be hard, if not
impossible, to get the entanglement limit by a prepare and measure
scheme. In our opinion, this is the main open question that
naturally follows from our analysis. The easiest way of
illustrating this problem is by considering the simple qubit
depolarizing channel of depolarizing probability $1-p$. This is a
channel where the input state is unchanged with probability $p$
and map into completely depolarized noise with probability
$1-p$. The corresponding state is a two-qubit Werner state. When
$p=1/3$, the channel is entanglement breaking, that is, it does
not allow to distribute entanglement, so it is useless for any
form of QKD. As shown here, the same channel can be used to QKD
using a prepare and measure scheme when $p>1/\sqrt 5$. Trivially,
the entanglement limit can be reached if one allows coherent
protocols by the two parties, such as entanglement distillation.
However, is there a prepare and measure scheme with positive key
rate for $1/3\leq p<1/\sqrt 5$?

\section{Acknowledgement}
We thank Emili Bagan, Cyril Branciard, Barbara Kraus, Llu\'\i s Masanes,
Ram\'on Mu\~noz-Tapia and Renato Renner for useful discussion.
This work is supported by the Spanish MEC, under a ``Ram\'on y
Cajal" grant and FIS 2004-05639 project, and the Generalitat de Catalunya, 2006FIR-000082 grant.

\section*{Appendix I. Cloning Based Attacks}

Asymmetric cloning machines have been proven to be a useful tool
in the study of optimal eavesdropping attacks. In a cryptographic
scenario, the input state to the cloning machine is the one sent
by Alice, while one of the outputs is forwarded by Eve to Bob,
keeping the rest of the output state. For instance, in the BB84
case, where Alice uses states from the $x$ and $z$ bases, the
optimal eavesdropping attack is done by a $1\rightarrow 1 + 1 $
phase-covariant cloning machine \cite{phase-covariant cloning}
that clones the $xz$ equator. The output states for Bob and Eve
are \bea \rho_{B} & = &
\frac{1}{2}(I+\eta_{xz}^{B}(n_{x}^{B}\sigma_{x}+n_{z}^{B}\sigma_{z})
+\eta_{y}^{B}n_{y}^{B}\sigma_{y}) \nonumber \\
\rho_{E} & = &
\frac{1}{2}(I+\eta_{xz}^{E}(n_{x}^{E}\sigma_{x}+n_{z}^{E}\sigma_{z})
+\eta_{y}^{E}n_{y}^{E}\sigma_{y}), \nonumber  \eea where $\eta_i$
are usually called the shrinking factors.

In the entanglement picture, this attack corresponds to the Bell
diagonal state \bea \rho_{AB} &=&
\lambda_{1}[\Phi_{1}]+\lambda[\Phi_{2}]
+\lambda[\Phi_{3}]+\lambda_{4}[\Phi_{4}].\nonumber  \eea Here
$\lambda_2=\lambda_3=\lambda$, which implies that the error rate
is the same in both bases. The normalization condition is
$\lambda_{1} + 2\lambda + \lambda_{4} =1$. When compared to the
cloning machine, the shrinking factor are $\eta_{xz}^{B} =
\lambda_{1} - \lambda_{4}$ and $\eta_{xz}^{E} =
2\sqrt{\lambda}(\sqrt{\lambda_{1}} + \sqrt{\lambda_{4}})$. Note
that $\eta_{y}^{B} = 1- 4 \lambda +4 \lambda_{4}$ and
$\eta_{y}^{E} = 2(\lambda +
\sqrt{\lambda_{4}(1-2\lambda-\lambda_{4})})$.

In the case of using one-way communication distillation protocols,
Eve's goal is to maximize, for a given QBER, her Holevo
information with Alice (see Eq. (\ref{dw})). The optimal
coefficients, or cloning attack, are $\lambda_{1}=(1-Q)^{2}$,
$\lambda = Q-Q^{2}$, and $\lambda_{4} = Q^{4}$, where $Q$ is the
QBER. When considering two-way communication protocols, as in this
work, the security condition is given in Sec. \ref{secsuffcond}.
According to this condition, the optimal coefficients are
$\lambda_{1}=1-2Q$, $\lambda = Q$, and $\lambda_{4} = 0$.

\section*{Appendix II. Eve's information in the case of pre-processing}

In this appendix, we show how to compute Eve's information in the
case Alice applies pre-processing before the CAD protocol, for large blocks. In this limit, Eve is faced with two
possibilities, $\rho_{0,0}^{\otimes N}$ and $\rho_{1,1}^{\otimes N}$, that read
\begin{eqnarray}
  \rho_{0,0} &=& u[e_{0,0}]+v[e_{0,1}] \nonumber\\
  \rho_{1,1} &=& u[e_{1,1}]+v[e_{1,0}]
\end{eqnarray}  Indeed, if $N\gg 1$, there are almost no errors in the symbols
accepted by Alice and Bob. Eve's Holevo bound then reads
\begin{equation}
    \chi(A:E)\approx S(\sigma_E)-Nh(u) ,
\end{equation}
where we used the fact that $S(\rho_{0,0}^{\otimes N})=S(\rho_{1,1}^{\otimes N})=Nh(u)$.

The main problem, then, consists of the diagonalization of
$\sigma_E$. Note however that the states $\rho_{0,0}$ and
$\rho_{1,1}$ have rank two and their eigenvectors belong to
different two-dimensional subspaces. This implies that $\sigma_E$
decomposes into two-dimensional subspaces that can be easily
diagonalized. The corresponding eigenvalues are
\begin{equation}
    \lambda_r=u^r v^{N-r}\frac{1\pm|\langle e_{0,0}\ket{e_{1,1}}|^r
    |\langle e_{0,1}\ket{e_{1,0}}|^{N-r}}{2}
\end{equation}
for $r=0,\ldots,N$, with degeneracy $N!/(r!(N-r)!)$. Replacing
these eigenvalues into the von Neumann entropy, one gets
\bea
    &&S(\sigma_E)=Nh(u)+\sum_{r=0}^N \begin{pmatrix}
      N \\
      r \\
    \end{pmatrix}u^r v^{N-r}\nonumber\\&&h\left
    (\frac{1+|\langle e_{0,0}\ket{e_{1,1}}|^r
    |\langle e_{0,1}\ket{e_{1,0}}|^{N-r}}{2}\right) .
\eea For large $N$ and nonzero $u$, the only relevant terms in the
previous sum are such that $|\langle e_{0,0}\ket{e_{1,1}}|^r
    |\langle e_{0,1}\ket{e_{1,0}}|^{N-r}\ll 1$. One can then
approximate $h((1+x)/2)\approx 1-x^2/\ln 4$, having
\begin{equation}
    S(\sigma_E)\approx Nh(u)+1-\frac{(u|\langle e_{0,0}\ket{e_{1,1}}|^2
    +v|\langle e_{0,1}\ket{e_{1,0}}|^2)^{N}}{\ln 4} ,
\end{equation}
where we used the binomial expansion. Collecting all the terms,
Eve's information reads
\begin{equation}
    \chi(A:E)\approx 1-\frac{(u|\langle e_{0,0}\ket{e_{1,1}}|^2
    +v|\langle e_{0,1}\ket{e_{1,0}}|^2)^{N}}{\ln 4} .
\end{equation}

\section*{ Appendix III. Properties of geometrically uniform states}

A set of $d$ quantum states $\{|\psi_{0}\rangle
,...,|\psi_{d-1}\rangle \}$ is said to be geometrically uniform if
there is a unitary operator $U$ that transforms $|\psi_{j}\rangle$
into $|\psi_{j+1}\rangle$ for all $j$, where the indices read
modulo $d$. All sets of geometrically uniform states, if the
cardinality is the same, are isomorphic. Therefore, we do not lose
any generality when assuming that those states are of the form:
\bea |\psi_{\alpha}\rangle = \sum_{n=0}^{d-1} c_{n} e^{\frac{2\pi
 i}{d}n\alpha} |x_{n}\rangle \nonumber \eea
where $\alpha$ runs from $0$ to $d-1$ and $|x_{n}\rangle$ are
orthonormal basis. Each state $|\psi_{\alpha}\rangle $ translates
to $|\psi_{\alpha+\beta}\rangle$ by applying $\beta$ times the
unitary $U = \sum_{m = 0}^{d-1}e^{\frac{2\pi
i}{d}m}|x_{m}\rangle\langle x_{m}|$. These states satisfy the
following properties, that are used in our computations:

\begin{itemize}
    \item Given a set of geometrically uniform states $\{|\psi_{0}\rangle
,...,|\psi_{d-1}\rangle \}$, an orthonormal basis spanning the
support of those states can
    explicitly obtained as follows: \bea |x_{n}\rangle =
\frac{1}{d c_{n}} \sum_{\alpha} e^{-\frac{2 \pi
i}{d}n\alpha}|\psi_{\alpha}\rangle. \label{basis} \eea

    \item The uniform mixture of geometrically uniform states
    gives the orthogonal decomposition in the basis defined above $\{|x_{n}\rangle
    \}$: \bea \rho &
= & \frac{1}{d} \sum_{\alpha} |\psi_{\alpha}\rangle \langle
\psi_{\alpha}| =\sum_{n} c_{n}^{2}|x_{n}\rangle \langle x_{n}|.
\nonumber \eea
\end{itemize}

Therefore, the eigenvalues of the equal mixture of geometrically
uniform state are $c_{n}^{2}$. Using (\ref{basis}), these
eigenvalues can be written as: \bea c_{n}^{2} = \frac{1}{d^{2}}
\sum_{\alpha,\beta} e^{\frac{2\pi i}{d}n(\beta - \alpha)} \langle
\psi_{\beta} |\psi_{\alpha}\rangle. \label{eig} \eea

In our case, we are interested in the eigenvalues of the state
\bea \rho = \frac{1}{d} \sum_{\alpha} |e_{\alpha}\rangle \langle
e_{\alpha}|^{\otimes N}, \nonumber \eea which approximates Eve's
state after CAD in the limit of large $N$. The states
$|e_{\alpha}\rangle^{\otimes N}$ are geometrically uniform, so the
searched eigenvalues are:

\bea \lambda_{\mu} = \frac{1}{d^{2}} \sum_{\alpha,\beta}
e^{\frac{2\pi i}{d}\mu(\beta - \alpha)} \langle e_{\beta}
|e_{\alpha}\rangle^{N}. \nonumber \eea

\section*{Appendix IV. Square-Root Measurement(SRM)}

We describe the so-called square-root measurement along the lines
given in Ref. \cite{srm}. Suppose that Alice encodes a classical
random variable $i$ that can take $l$ different values into a
quantum state $|\phi_{i}\rangle\in\compl^d$, with $l\leq d$, and
sends the state to Bob. Suppose the $l$ states are non-orthogonal
and span an $m$ dimensional subspace of $\compl^d$. Denote by
$\Pi_m$ the projection into this subspace, i.e.
$\Pi_m\ket{\phi_i}=\ket{\phi_i}$ for all $i$. Bob has to read out
the encoded value from the quantum state in an ``optimal" way.
There exist several ``optimal" measurements depending on the
figure of merit to be optimized. Here, following \cite{srm}, we
consider that Bob applies a measurement consisting of $l$ rank-one
operators $[m_{i}]$, satisfying $\sum_i[m_i]=\Pi_m$. The figure of
merit to be optimized is the squared error $E =
\sum_{i=0}^{l-1}\langle E_{i}|E_{i}\rangle$, where $|E_{i}\rangle
= |\phi_{i}\rangle - |m_{i}\rangle$ are the error vectors. As
shown in \cite{srm}, the measurement strategy minimizing $E$ is
the so-called SRM, also known as pretty-good
measurement. The construction of this optimal measurement works as
follows.

Denoted by $\Phi$ the matrix whose columns are $|\phi_{i}\rangle$.
The SRM is constructed from the structure of the matrix $\Phi$.
Applying singular value decomposition to $\Phi = U D V^{\dagger}$,
the optimal measurement matrix is \cite{srm} \bea M = \sum_{i}
|u_{i}\rangle \langle v_{i}| \nonumber \eea where $|u_{i}\rangle$
and $|v_{i}\rangle$ are the column vectors of the two unitary
matrices $U$ and $V$, respectively. Here the column vectors of $M$
define the optimal choice of measurement projectors $\ket{m_i}$.

Moving to our cryptography problem, the states Eve has to
discriminate are the geometrically uniform states \bea
|e_{\gamma}\rangle = \sum_{n=0}^{d-1}\beta_{n} e^{2\pi i (\gamma n
/d)} |x_{n}\rangle \nonumber \eea where $|x_{n}\rangle$ is an
orthonormal basis in a $d$-dimensional Hilbert space, and $\gamma
$ runs from $0$ to $d-1$. Each $|e_{\gamma}\rangle$ is normalized.
In our problem, Eve aims at minimizing her error probability.
Interestingly, in the case of geometrically uniform state, the
previous measurement strategy turns out to minimize the error
probability as well \cite{srm}. So, we only have to derive the
optimal measurement matrix from $\Phi = \sum_{\gamma}
|e_{\gamma}\rangle \langle x_{\gamma}|$. Using relations
$\Phi^{\dagger}\Phi = V D V^{\dagger}$, the unitary $V$ is the
$d$-dimensional Fourier transform $\mathcal{F}|x_{u}\rangle =
\frac{1}{\sqrt{d}}\sum_{w} \exp (-\frac{2\pi
i}{d}wu)|x_{w}\rangle$, and the diagonal matrix is
$D=diag(\sqrt{d}|\beta_{n}|)$. Therefore, the optimal measurement
matrix is \bea M = \sum_{i} |m_{i}\rangle \langle x_{i}| \nonumber
\eea where \bea |m_{j}\rangle  =
\frac{1}{\sqrt{d}}\sum_{k=0}^{d-1}e^{\frac{2\pi
i}{d}jk}|x_{k}\rangle \nonumber \eea

Using this measurement, the probability of guessing correctly a
given state $|e_{j}\rangle $ is $|\langle m_{j} | e_{j}\rangle
|^{2}$. Then, the average success probability is \be P^{success}
=  \sum_{j=0}^{d-1} p(j)\, |\langle m_{j} | e_{j}\rangle |^{2} =
\frac{1}{d}\big|\sum_{n}\beta_{n}\big|^{2} \label{measurement}\ee
The last equality is obtained taking into account that all
$|e_{j}\rangle$ are equally probable, $p(j) = 1/d$. In particular,
for  the $d+1$- or 2-bases protocols, the success probability
reads, in terms of $v$ and $z$, $P^{success} = (v+(d-1)z)^{2}/dF$.

When $N$ copies of the states are given, $|e_{j}\rangle^{\otimes
N}$, we can apply a collective measurement strategy. The SRM is
constructed in the same way as above, and the success probability,
assuming that all states are equi-probable, is \be
P_{N}^{success} = \frac{1}{d^{2}}\left|~\sum_{\eta}
\sqrt{\sum_{m} e^{2\pi i (\eta m/d)}
\langle e_{m}|e_{0}\rangle^{N} }~\right|^{2} .
\ee



\begin{thebibliography}{99}
\bibitem{notation CCQ}
Throughout this paper, we denote classical and quantum variables by
C and Q, respectively. When writing correlations among the three parties, the
order is Alice-Bob-Eve. For instance, CCQ means that Alice and Bob
have correlated classical values (for instance, after some measurements), while Eve has a quantum state.
\bibitem{note prepr}
It is worth mentioning here that some of the techniques studied in this section may improve the key rate for some values of the error rate.
However, we prove here that they do not improve the critical
tollerable error rate.
\bibitem{notemub}
The existence of the $d+1$ mutually unbiased bases in nay
dimension is a well-known open problem. This existence has only
been proven in the case where the dimension is a power of a prime
number. Recall that two bases, $\{\ket{u_i}\}$ and $\{\ket{v_i}\}$
with $i=1,\ldots,d$, are said mutually unbiased whenever $|\langle
u_i\ket{v_j}|^2=1/d$.
\end{thebibliography}
\end{document}